\documentclass[prd,aps,showpacs,11pt,
nofootinbib,%
tightenlines,
notitlepage,
]{revtex4-1}

\usepackage{graphicx}
\usepackage{amsmath}
\usepackage{amssymb}
\usepackage{hyperref} 

\newcommand{\be}{\begin{equation}}
\newcommand{\ee}{\end{equation}}
\newcommand{\bea}{\begin{eqnarray}}
\newcommand{\eea}{\end{eqnarray}}

\begin{document}

\title{Hadronic light-by-light contribution to 
the muon $g-2$\\
from holographic QCD with massive pions}

\author{Josef Leutgeb}
\author{Anton Rebhan}
\affiliation{Institut f\"ur Theoretische Physik, Technische Universit\"at Wien,
        Wiedner Hauptstrasse 8-10, A-1040 Vienna, Austria}

\date{\today}

\begin{abstract}
We extend our previous calculations of the hadronic light-by-light scattering contribution to
the muon anomalous magnetic moment in holographic QCD to models with finite quark masses and a tower of massive pions. Analysing the role of the latter in the
Wess-Zumino-Witten action,
we show that the Melnikov-Vainshtein short-distance constraint is satisfied solely by
the summation of contributions from the infinite tower of axial vector meson contributions.
There is also an enhancement of the asymptotic behavior of pseudoscalar contributions when their
infinite tower of excitations is summed, but this 
leads only to subleading contributions for the
short-distance constraints on light-by-light scattering.
We also refine our numerical evaluations, particularly in the pion and $a_1$ sector,
which corroborates our previous findings of contributions from axial vector
mesons that are significantly larger than those adopted for the effects of
axials and short-distance constraints in the recent White Paper
on the Standard Model prediction for $(g-2)_\mu$.
\end{abstract}

\maketitle

\section{Introduction}

Recently, the Muon $g-2$ Collaboration at Fermilab
\cite{Muong-2:2021ojo}
has confirmed the long-standing discrepancy between
the E821/BNL measurement \cite{Muong-2:2006rrc} of the anomalous magnetic moment of the
muon \cite{Jegerlehner:2017gek} and the Standard Model prediction, which according to the White Paper (WP) of the Muon $g-2$ Theory Initiative \cite{Aoyama:2020ynm} is
below the combined experimental value by an amount of $\Delta a_\mu=\Delta(g-2)_\mu/2 = 251(59)\times10^{-11}$.

While
QED \cite{Aoyama:2012wk,*Aoyama:2017uqe,*Aoyama:2019ryr} and electroweak effects \cite{Czarnecki:2002nt,Gnendiger:2013pva} appear
to be fully under control, the theoretical uncertainty
is dominated by hadronic effects
\cite{Melnikov:2003xd,Prades:2009tw,Kurz:2014wya,Colangelo:2014qya,Pauk:2014rta,Davier:2017zfy,Masjuan:2017tvw,Colangelo:2017fiz,Keshavarzi:2018mgv,Colangelo:2018mtw,Hoferichter:2019gzf,Davier:2019can,Keshavarzi:2019abf,Hoferichter:2018dmo,*Hoferichter:2018kwz,Gerardin:2019vio,Bijnens:2019ghy,Colangelo:2019lpu,Colangelo:2019uex,Danilkin:2019mhd,Blum:2019ugy,Chao:2021tvp}.
The largest contribution by far is the hadronic vacuum polarization,
where a recent lattice calculation \cite{Borsanyi:2020mff}
has challenged the WP result, albeit by producing tensions in other
quantities \cite{Crivellin:2020zul,Keshavarzi:2020bfy,Colangelo:2020lcg} --
an issue which further lattice
calculations as well as new experimental results on hadronic cross
sections should resolve in the near future.
The second largest uncertainty comes from
hadronic light-by-light scattering (HLBL) \cite{Danilkin:2019mhd}, which also needs
improvements in view of the expected reductions of experimental
errors by the ongoing experiment at Fermilab and a new one in preparation
at J-PARC \cite{Abe:2019thb}.

At the low energies relevant for the evaluation of $a_\mu$, the HLBL
amplitude is dominated by the two-photon coupling of the pion, where
the transition form factors (TFF) needed to evaluate the pion pole
contribution has been computed with satisfying accuracy in
data-driven approaches \cite{Masjuan:2017tvw,Hoferichter:2018dmo,*Hoferichter:2018kwz} which are also
in agreement with results from 
lattice QCD \cite{Gerardin:2019vio}.
Among the many additional low-energy contributions, those
contributed by axial vector mesons are currently the ones
involving the largest uncertainties due to insufficient experimental
data. Axial vector mesons also play a crucial role in connecting
the HLBL amplitude at low energies to short-distance constraints (SDCs),
in particular the one derived by Melnikov and Vainshtein (MV) \cite{Melnikov:2003xd} using
the non-renormalization theorem for the vector-vector-axial-vector (VVA)
correlator and the axial anomaly, as has been clarified by holographic models of QCD (hQCD) \cite{Leutgeb:2019gbz,Cappiello:2019hwh}, which are the first hadronic models to implement fully
the constraints \cite{Knecht:2020xyr,Masjuan:2020jsf} imposed by the axial anomaly in the chiral limit.

Away from the chiral limit, also excited pseudoscalars couple to the axial
vector current and thus to the axial anomaly. In \cite{Colangelo:2019lpu,Colangelo:2019uex} a Regge model for excited pseudoscalars
has been constructed to satisfy the longitudinal SDCs (LSDCs) and to estimate
the corresponding contributions to $a_\mu$.
In this paper we employ the simplest hard-wall AdS/QCD model that is
capable of satisfying at the same time the leading-order (LO) perturbative
QCD (pQCD) constraints on the vector correlator and the low-energy
parameters provided by the pion decay constant $f_\pi$ and the mass of the
$\rho$ meson, introduced in \cite{Erlich:2005qh,DaRold:2005mxj,*DaRold:2005vr} and
called HW1 in \cite{Cappiello:2010uy,Leutgeb:2019zpq,Leutgeb:2019gbz},
using here its generic form with finite quark masses (albeit in the simplest
version of uniform quark masses).
This allows us to study the role of massive pions in the axial anomaly
and in the longitudinal SDCs. We find that in hQCD
there is in fact a certain enhancement of the 
asymptotic behavior when summing the
infinite tower of pseudoscalars compared to the behavior of individual contributions,
but within the allowed range of parameters of hQCD this is insufficient
to let pseudoscalars contribute to the leading terms of the longitudinal SDCs; the latter are
determined by the infinite tower of axial vector mesons alone,
in agreement with the 
expectation expressed in \cite{Cappiello:2019hwh}.

Moreover, we evaluate and assess a set of massive hard-wall AdS/QCD models with regard
to their predictions for the pseudoscalar and axial-vector TFFs
and the resulting contributions to $a_\mu$, also allowing for
adjustments at high energies to account for next-to-leading (NLO)
pQCD corrections. 
The results for the single and double-virtual pion TFFs agree perfectly with those of the data-driven
dispersive results of Ref.~\cite{Hoferichter:2018kwz}, in particular when
a 10\% reduction of the asymptotic limit is applied.
The shape of the axial vector TFF is consistent with L3 results for the $f_1(1285)$
axial vector meson \cite{Achard:2001uu}; the magnitude is above experimental
values, but becomes compatible after such reductions. 
While hQCD is clearly
only a toy model for real QCD, this success after a fit of a minimal set of
low-energy parameters seems to make it also interesting as a phenomenological model
(which may be further improved by using a modified background geometry
and other refinements \cite{Ghoroku:2005vt,Karch:2006pv,Gursoy:2007cb,*Gursoy:2007er}).

Numerically, the axial vector contributions are close to our previous
results for the HW1 model in the chiral limit (they even tend to increase
with finite quark masses), and they are thus significantly larger than
estimated in the WP \cite{Aoyama:2020ynm}. 
Interestingly enough, a new
complete lattice calculation 
which claims comparable errors \cite{Chao:2021tvp}
has obtained somewhat larger values of $a_\mu^\mathrm{HLBL}$ than the WP \cite{Aoyama:2020ynm},
which could also be indicative of larger contributions from the axial vector meson sector.

The contributions from the excited pseudoscalars are numerically
much smaller than those from axial vector mesons. They are present also
in the chiral HW1 model, where they decouple from the axial vector current
and the axial anomaly, but not from low-energy photons. However, the first excited
pion mode has a two-photon coupling that is significantly above the experimental constraint deduced in \cite{Colangelo:2019uex}. Nevertheless, the contribution
of the first few excited modes together is smaller than those
of the Regge model of \cite{Colangelo:2019lpu,Colangelo:2019uex}, which
is compatible with known experimental constraints. This model was, however,
not meant primarily as a phenomenological model for estimating the
contributions of excited pseudoscalars but rather as a model for
estimating the effects of the longitudinal SDCs, which according to the hQCD models
are instead provided by the axial vector mesons.
As far as these effects are concerned,
our present results are 
not far above the estimates obtained in
\cite{Colangelo:2019lpu,Colangelo:2019uex,Ludtke:2020moa,Colangelo:2021nkr};
all are significantly below the estimates obtained with the so-called MV model \cite{Melnikov:2003xd}, where the ground-state pion contribution is modified.

This paper is organized as follows. In Sec.~\ref{sec:HW} we review
hQCD with an anti-de Sitter (AdS) background that is cut off by a hard wall
and where in addition to flavor gauge fields a bifundamental scalar bulk field
encodes the chiral condensates with or without finite quark masses.
In addition to the boundary conditions considered originally in Ref.~\cite{Erlich:2005qh,DaRold:2005mxj} where vector and axial vector gauge
fields are treated equally (HW1), we also consider the other
set of admissible boundary conditions that appear in models
without a bifundamental scalar, the Hirn-Sanz (HW2) model \cite{Hirn:2005nr}
and the top-down hQCD model of Sakai and Sugimoto \cite{Sakai:2004cn,Sakai:2005yt}.
(This modification of the HW1 model
will be referred to as HW3.)
As has been pointed out in \cite{Domenech:2010aq}, this has the
advantage of removing infrared boundary terms in the Chern-Simons action
that in the HW1 model need to be subtracted by hand \cite{Grigoryan:2007wn}.
Additionally we consider the generalization of different scaling dimensions
for quark masses and chiral condensates proposed in \cite{Domenech:2010aq},
which permits to fit the mass of the first excited pion or the lowest
axial vector meson.
In Sec.~\ref{sec:CS}, we analyse the consequences of the axial anomaly
in the HW models, in particular the role of excited pseudoscalars, as encoded
by the bulk Chern-Simons action for the five-dimensional gauge fields.
In Sec.~\ref{sec:SDC} we study SDCs on TFFs and the HLBL scattering amplitude,
showing analytically that the MV-SDC is always saturated by axial vector mesons,
while the symmetric longitudinal SDC is fulfilled to 81\%.
Within the bounds of allowed scaling dimensions for the bulk bifundamental scalar,
the infinite tower of excited pseudoscalar cannot change this.
Finally in Sec.~\ref{sec:results} we evaluate the set of massive HW models
numerically, comparing the resulting masses, decay constants, and TFFs with
empirical data, before calculating the corresponding contributions to $a_\mu$.

\section{Hard-wall AdS/QCD models with finite quark masses}\label{sec:HW}

In the hard-wall AdS/QCD models of Ref.~\cite{Erlich:2005qh,DaRold:2005mxj,Hirn:2005nr}, the background geometry
is chosen as pure anti-de Sitter 
with metric
\be
ds^2=z^{-2}(\eta_{\mu\nu}dx^\mu dx^\nu - dz^2),
\ee
with a conformal boundary at $z=0$. Confinement is implemented by
a cutoff at some finite value of the radial coordinate $z_0$,
where boundary conditions for the five-dimensional fields are imposed.

Left and right chiral quark currents are dual to five-dimensional flavor gauge fields, whose action
is given by
a Yang-Mills part\footnote{Note that in our conventions the metric $g_{MN}$ carries a dimension
of inverse length squared; $g_5$ is therefore dimensionless.}
\bea
S_{\rm YM} = -\frac{1}{4g_5^2} \int d^4x \int_0^{z_0} dz\,
\sqrt{-g}\, g^{PR}g^{QS}
\,\text{tr}\left(\mathcal{F}^\mathrm{L}_{PQ}\mathcal{F}^\mathrm{L}_{RS}
+\mathcal{F}^\mathrm{R}_{PQ}\mathcal{F}^\mathrm{R}_{RS}\right),
\eea
where $P,Q,R,S=0,\dots,3,z$ and $\mathcal{F}_{MN}=\partial_M \mathcal{B}_N-\partial_N \mathcal{B}_M-i[\mathcal{B}_M,\mathcal{B}_N]$, and
a Chern-Simons action $S_{\rm CS}=S_{\rm CS}^\mathrm{L}-S_{\rm CS}^\mathrm{R}$ with (in
differential form notation)
\be\label{SCS}
S_{\rm CS}^\mathrm{L,R}=\frac{N_c}{24\pi^2}\int\text{tr}\left(\mathcal{B}\mathcal{F}^2-\frac{i}2 \mathcal{B}^3\mathcal{F}
-\frac1{10}\mathcal{B}^5\right)^\mathrm{L,R},
\ee
up to a potential subtraction of boundary terms at $z_0$ to be discussed below.

As dual field for the scalar quark bilinear operator, 
a bi-fundamental bulk scalar \cite{Erlich:2005qh,DaRold:2005mxj} $X$,
parametrized as
\cite{Abidin:2009aj}
\be
X=e^{i\pi^a(x,z) t^a}[{\textstyle\frac12} v(z)
]e^{i\pi^a(x,z)t^a},
\ee
is introduced, with action

\be
S_X=\int d^4x \int_0^{z_0} dz\,\sqrt{-g}\,\text{tr}\left(
|DX|^2-M_X^2|X|^2 \right),
\ee
where $D_M X=\partial_M X-i \mathcal{B}_M^\mathrm{L} X+iX\mathcal{B}_M^\mathrm{R}$.
The five-dimensional mass term is determined by the scaling dimension 
of the chiral-symmetry
breaking order parameter $\bar q_L q_R$ of the boundary theory.
With $M_X^2=-3$, one obtains
vacuum solutions
$v(z)=M_q z+\Sigma\, z^3$, 
where $M_q$ and $\Sigma$ are
model parameters related to the quark mass matrix and the quark condensate.\footnote{According to \cite{DaRold:2005vr,Cherman:2008eh,Gorsky:2009ma,Abidin:2009aj} these need to be rescaled
by $C=\sqrt{N_c}/(2\pi)$ before being interpreted as quark masses and
condensates: $m_q=M_q/C$ and $\langle q \bar q\rangle=C\Sigma$.\label{footnote:rescaling}}
Following Ref.~\cite{Domenech:2010aq} we shall also consider generalizations
away from $M_X^2=-3$, where 
\be
v(z)=M_q z_0 (z/z_0)^{\Delta^-}
+\Sigma\, z_0^3 (z/z_0)^{\Delta^+}
\ee
with 
\be
\Delta^\pm=2\pm\sqrt{4 + M_X^2}\equiv 2\pm\alpha.
\ee

In the original hard-wall model of Ref.~\cite{Erlich:2005qh,DaRold:2005mxj} (HW1),
chiral symmetry preserving boundary conditions are employed 
at the infrared cutoff, $\mathcal{F}^\mathrm{L,R}_{z\mu}(z_0)=0$,
whereas in the model of Hirn and Sanz \cite{Hirn:2005nr} (called HW2 in \cite{Cappiello:2010uy,Leutgeb:2019zpq,Leutgeb:2019gbz})
chiral symmetry is broken  through $(\mathcal{B}^\mathrm{L}_\mu-\mathcal{B}^\mathrm{R}_\mu)(z_0)=0$
and $(\mathcal{F}^\mathrm{L}_{z\mu}+\mathcal{F}^\mathrm{R}_{z\mu})(z_0)=0$
without the introduction of a bi-fundamental
scalar.
These latter conditions arise naturally in the top-down holographic
chiral model of Sakai and Sugimoto \cite{Sakai:2004cn,Sakai:2005yt} with flavor gauge fields residing on
flavor branes separated by an extra dimension at the boundary but connecting in the bulk. In \cite{Domenech:2010aq}, it was proposed to use such
symmetry breaking boundary conditions also in the presence of
symmetry breaking by a bifundamental scalar, because it avoids infrared
boundary contributions from the CS action. This variant of the HW1 model
will be termed HW3 in the following.

In both, the HW1 and the HW3 model, chiral symmetry can be broken additionally by finite quark masses. In the present paper, we shall only consider the
flavor symmetric case, where both $M_q$ and $\Sigma$ are proportional to
the unit matrix. For both models we will also consider generalizations away from $M_X^2=-3$,
which, as Ref.~\cite{Domenech:2010aq} has found, allows the HW3 model
to fit simultaneously the masses of the lightest and the first excited pion. As shown in Appendix \ref{App:GOR}, all these models
lead to a Gell-Mann--Oakes--Renner (GOR) relation in the limit of small quark masses, to wit,
\be
f_\pi^2 m_\pi^2=
2\alpha 
M_q \Sigma \quad\mbox{for}\;
M_q \ll \Sigma z_0^2\propto \Sigma/m_\rho^2,
\ee
where $\alpha
=1$ for the standard choice $M_X^2=-3$, and
$0<\alpha<2$ for the admissible generalizations \cite{Domenech:2010aq}
$-4<M_X^2<0$.

\subsection{Vector sector}

As long as flavor symmetry is in place, vector mesons, which
appear in the mode expansion of $V_\mu=(\mathcal{B}^L_\mu
+\mathcal{B}^R_\mu)/2$,
have (for all hard-wall models considered here) quark-mass
independent holographic wave functions given by 
\be\label{psinHW}
\partial_z\left[\frac1z \partial_z \psi_n(z)\right]+\frac1z M_n^2 \psi_n(z)=0
\ee
with boundary conditions $\psi_n(0)=\psi'_n(z_0)=0$.
This is solved by $\psi_n(z)\propto zJ_1(M_n z)$ with $M_n$ determined by
the zeros of the Bessel function $J_0$, denoted by $\gamma_{0,n}$. 
With normalization $\int_0^{z_0} dz\,z^{-1}\psi_n(z)^2=1$ we have
\be
\psi_n(z)=\sqrt{2}\frac{z J_1(\gamma_{0,n}z/z_0)}{z_0J_1(\gamma_{0,n})}
\ee
and canonically normalized vector meson fields $v_\mu(x)$ in the mode expansion
$V_\mu^a=g_5\sum_{n=1}^\infty v_\mu^a(x)\psi_n(z)$.

Identifying $M_1=\gamma_{0,1}/z_0=m_\rho=775$ MeV, we obtain
\be\label{eq:z0}
z_0=\gamma_{0,1}/m_\rho=3.103 \, \text{GeV}^{-1}.
\ee

The vector bulk-to-boundary propagator is obtained by replacing $M_n^2$
by $q^2=-Q^2$ and the boundary conditions by
$\mathcal{J}(Q,0)=1$ and $\partial_z \mathcal{J}(Q,z_0)=0$. This gives
\be\label{HWVF}
\mathcal{J}(Q,z)=
Qz \left[ K_1(Qz)+\frac{K_0(Q z_0)}{I_0(Q z_0)}I_1(Q z) \right]=
\sum_{n=1}^\infty \frac{g_5 F_n^V \psi_n(z)}{Q^2+M_n^2},
\ee
where $F_n^V=|\psi_n'(\epsilon)/(g_5\epsilon)|$ is the decay constant of $v_n(x)$,
\be\label{FVdef}
\langle 0 | J_\mu^a(0) | v_n^b(q,\lambda) \rangle = F_n^V \varepsilon_\mu(q,\lambda) \delta^{ab}.
\ee

Requiring that the vector current two-point function matches
the leading-order pQCD result \cite{Erlich:2005qh},
\be\label{PiVas}
\Pi_V(Q^2)=-\frac1{g_5^2 Q^2} \left( \frac1z \partial_z \mathcal{J}(Q,z) \right)\Big|_{z\to0}=-\frac{N_c}{24\pi^2}\ln Q^2,
\ee
determines $g_5^2=12\pi^2/N_c$.

The simpler Hirn-Sanz (HW2) model \cite{Hirn:2005nr}, 
which does not include the scalar $X$,
cannot set $g_5^2$ to the value required by pQCD, when the mass of
the rho meson and the pion decay constant is matched to experiment
(as we shall do throughout).

\subsection{Axial sector}

Following the notation of Ref.~\cite{Abidin:2009aj},
the five-dimensional Lagrangian contains the following quadratic terms
for the axial gauge field $A_M=(\mathcal{B}^L_M
-\mathcal{B}^R_M)/2$ and the pseudoscalar $\pi$,
\begin{equation}\label{Laxial}
\mathcal{L}_{axial}=-\frac{1}{4g_5^2 z}(\partial_M A_N -\partial_N A_M )^2+\frac{\beta(z)}{2 g_5^2 z} (\partial_M \pi - A_M )^2,
\qquad \beta(z)=g_5^2 v^2/z^2,
\end{equation}
where five-dimensional indices are implicitly contracted with
a five-dimensional Minkowski metric $\eta^{MN}$ (mostly minus). 

Axial vector mesons are contained in $A_\mu^\perp=g_5\sum_{n=1}^\infty a_\mu^{(n)}(x) \psi_n^A(z)$
with the holographic wave functions subject to
\be\label{psiAnHW}
\partial_z\left[\frac1z \partial_z \psi^A_n(z)\right]+\frac1z \left[M_{A,n}^2-\beta(z)\right] \psi^A_n(z)=0,
\ee
which is different than (\ref{psinHW}) when $v\not=0$ and so gives
a different mass spectrum even when vector and axial vector fields
have identical boundary conditions, as is the case in the HW1 model.
In the HW3 model, chiral symmetry is additionally broken by the
Dirichlet boundary condition $\psi_n^{A}(z_0)_\mathrm{HW3}=0$.

In the holographic gauge $A_z=0$, the pseudoscalar $\pi$ mixes with the longitudinal part of the axial vector gauge field denoted by $A_\mu^\parallel = \partial_\mu \phi$,
\begin{equation}
\mathcal{L}_{axial} \supset  \frac{1}{2 g_5^2 z} (\partial_\mu \partial_z \phi)^2 + \frac{1}{2 g_5^2 z} \beta(z) (\partial_\mu \pi - \partial_\mu \phi)^{2} - \frac{1}{2g_5^2 z} \beta(z) (\partial_{z} \pi)^2. \label{eq:LAxial}
\end{equation}
Alternatively, one may decouple $A_\mu$ from $A_z$ and $\pi$ by adding the $R_\xi$ gauge fixing term \cite{Domenech:2010aq}
\be
\mathcal{L}_{R_\xi}=-\frac1{2\xi z g_5^2}\left[
\partial_\mu A^\mu-\xi z \partial_z \left( \frac1z A_z\right)
+\xi \beta \pi \right]^2.
\ee
Taking the $\xi\to\infty$ limit imposes the unitary gauge,
which fixes
\be
z \partial_z \left( \frac1z A_z^\mathrm{U}\right)=\beta \pi^\mathrm{U}.
\ee
In the unitary gauge, $A_\mu^\mathrm{U}$ is a simple Proca field, and the
relation to the fields in the holographic gauge (the latter written without label) are
$A_M^\mathrm{U}=A_M-\partial_M\phi$, $\pi^\mathrm{U}=\pi-\phi$ with
$A_\mu^{\parallel\mathrm{U}}=0$ and $A_z^{\mathrm{U}}=-\partial_z\phi$.

The equations of motion for $\phi$ and $\pi$ read
\bea
\label{eq:phieom-1}
\partial_{z}\left(\frac{1}{z}\partial_{z}\phi\right)+\frac{\beta(z)}{z}\left(\pi-\phi\right) &=&0,\\
\partial_{z}\left(\frac{\beta(z)}{z}\partial_{z}\pi\right)+\frac{\beta(z)}{z}q^{2}\left(\pi-\phi\right)=0.\label{eq:pieom-1}
\eea

Defining $y=\partial_{z}\phi/z$, one can combine these coupled equations
into a single second order differential equation \cite{Kwee:2007dd,Abidin:2009aj}
\begin{equation}
(L+q^2)y:=\frac{\beta(z)}{z}\partial_{z}\left(\frac{z}{\beta(z)}\partial_{z}y\right)-\beta(z)y
+q^2y=0.
\label{eq:eomy}
\end{equation}

The modes for $\pi$ follow from (\ref{eq:phieom-1}) and (\ref{eq:pieom-1})
which imply
\begin{equation}\label{eq:eomdy}
\partial_{z}y  =\frac{\beta(z)}{z}\left(\phi-\pi\right)
\end{equation}
and
\begin{equation}\label{eq:eomdpi}
\partial_{z}\pi  =\frac{q^{2}}{\beta(z)}zy.
\end{equation}

The boundary conditions at $z=\epsilon\to0$
are
$\phi_{S}(q,\epsilon)=0$, $\pi_{S}(q,\epsilon)=-1$ for the non-normalizable profile functions, and $\phi_{n}(\epsilon)=0$, $\pi_{n}(\epsilon)=0$ for the normalizable modes. This implies
\be
\left.\frac{z}{\beta(z)}\partial_{z}y_{S}(q,z)\right|_{z=\epsilon}=1,
\quad
\left.\frac{z}{\beta(z)}\partial_{z}y_{n}(z)\right|_{z=\epsilon}=0.
\ee
At the infrared boundary $z=z_0$ there are two possible choices which
eliminate the boundary term in the axial part of the action
\be\label{eq:bdyterm}
\frac{1}{2g_5^2z_0}\left(q^2\phi\partial_z\phi-\beta\pi\partial_z\pi \right)_{z=z_0}=\frac{q^2}{2g_5^2}y(z_0)\left[\phi(z_0)-\pi(z_0)\right].
\ee
In the HW1 model, one has Neumann boundary conditions at $z=z_0$
for all gauge fields, and therefore on $\phi$; (\ref{eq:bdyterm}) is then
cancelled by requiring $y(z_0)=0$. In the HW3 model, where the
axial gauge field has Dirichlet boundary conditions, one chooses
instead $\partial_z y|_{z=z_0}=0$, which through (\ref{eq:eomdy})
implies $[\phi(z_0)-\pi(z_0)]=0$. 
The HW2/HW3 boundary condition $A_\mu(z_0)=0$ thus only applies to $A^\perp_\mu(z_0)=0$
when the holographic gauge $A_z=0$ is employed, where $A_\mu^\parallel=\partial_\mu\phi$;
in the unitary gauge the latter is transferred to $A_z^\mathrm{U}=-\partial_z\phi=-z y$.

\begin{figure}[h]
\centering
\includegraphics[width=0.475\textwidth]{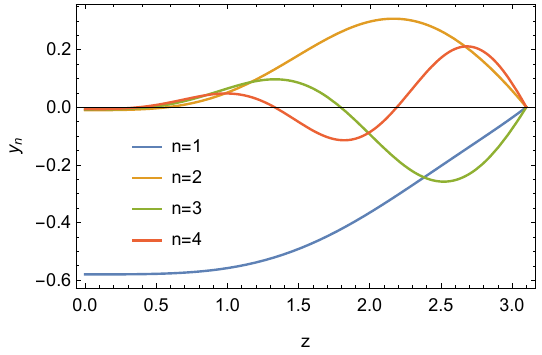}\qquad
\includegraphics[width=0.475\textwidth]{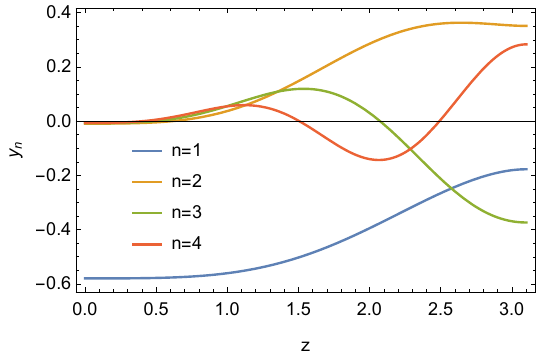}
\caption{The first four pion mode functions $y_n(z)$ in the HW1m model and in the HW3m version (precise definitions in Sec.~\ref{sec:results}), with $z$ in units of inverse GeV; the $y_n$ have units of GeV, with $y_n(0)=-g_5 f_{\pi_n}$.\label{fig:ymodes}}
\end{figure}

The normalizable modes are normalized by
\begin{equation}
\int_{\epsilon}^{z_{0}}dz\frac{z}{\beta(z)}y_{n}y_{m}= \int_{\epsilon}^{z_{0}}dz\frac{1}{z\beta(z)}\partial_{z}\phi_{n}\partial_{z}\phi_{m}= \frac{\delta_{mn}}{m_{n}^{2}}\label{eq:orthonoy}
\end{equation}
so that canonically normalized pion fields $\Pi_n(x)$ appear in a mode expansion
according to
\be
\phi=g_5 \sum_{n=1}^\infty \Pi_n(x)\phi_n(z),\quad 
\pi=g_5 \sum_{n=1}^\infty \Pi_n(x)\pi_n(z).
\ee
In unitary gauge one has 
\be\label{AUmodes}
A_z^\mathrm{U}=-\partial_z\phi=
-g_5 \sum_{n=1}^\infty \Pi_n(x) zy_n(z).
\ee

Normalizable and non-normalizable $y$-modes are related by the sum rule \cite{Abidin:2009aj}
\begin{equation}
y_{S}(q,z) =\sum_{n}\frac{m_{n}^{2}y_{n}(\epsilon)y_{n}(z)}{q^{2}-m_{n}^{2}}
=-g_5 \sum_{n=1}^\infty \frac{f_{\pi_n} m_n^2 y_n(z)}{q^2-m_n^2},
\label{sumrule}
\end{equation}
where $m_n$ are the masses of the tower of pions and $f_{\pi_n}=-y_n(\epsilon)/g_5$
their decay constants,
\be
\langle 0 | J^{A,a}_{\parallel\mu}(0)|\pi_n^b(q)\rangle
=i f_{\pi_n} q_\mu \delta^{ab}.
\ee

For later use we also define the Green function
\begin{equation}
\frac{z}{\beta(z)}\left(L+q^{2}\right)G(z,z^{\prime};q)= -\delta(z-z^{\prime})\label{Greendef}
\end{equation}
with boundary conditions as imposed on the mode functions $y_n$, so that
\be
G(z,z^{\prime};q)=  -\sum_{n}\frac{m_{n}^{2}y_{n}(z)y_{n}(z^{\prime})}{q^{2}-m_{n}^{2}}.\label{Gsumrule}
\ee

In the chiral limit, the infinite tower of massive pions continues to be present, but their decay constants vanish: $f_{\pi_n}\to0$ for $n\ge2$
as $M_q\to0$. The mode $n=1$ becomes massless with $y_S\to g_5 f_\pi y_1$, $f_\pi=f_{\pi_1}$. (See Appendix \ref{App:GOR} for more details.)

\section{Axial anomaly and massive pions}\label{sec:CS}

The Chern-Simons action (\ref{SCS}) implements the axial anomaly and the associated
$VVA$ coupling \cite{Grigoryan:2007wn,Son:2010vc}. After integration over the holographic direction, one
obtains a Wess-Zumino-Witten action for the mesons. This contains vertices involving photons, which are described by
$V_\mu(x,z)=A_\mu^{e.m.}(x)\mathcal{Q}\mathcal{J}(z)$, with $\mathcal{Q}$
denoting the charge matrix of quarks, and the fields encoded by $A_M$,
namely the infinite towers of pions and axial vector mesons.

In the chiral limit, where all massive pions decouple from the axial vector current through vanishing
decay constants, the resulting pion TFF
has been analysed in various holographic QCD models in
\cite{Grigoryan:2008up,Grigoryan:2008cc} and also in 
\cite{Cappiello:2010uy,Leutgeb:2019zpq}, where the HLBL
contribution to the anomalous magnetic moment of the muon 
was studied. This was also done
in \cite{Hong:2009zw} for
the ground-state pseudo-Goldstone bosons in a version of the HW1 model with finite quark masses.

The contribution of the infinite tower of axial vector mesons,
which is crucial for satisfying the Melnikov-Vainshtein short-distance
constraint, has been worked out in \cite{Leutgeb:2019gbz,Cappiello:2019hwh} for 
the chiral version of the HW1 model
and the inherently chiral Hirn-Sanz (HW2) model. We refer to these latter
references for details on the axial vector contributions, which as we shall see do not
change qualitatively away from the chiral limit, concentrating here on
the generalizations necessary to include the contributions
of massive pions in the HW1 and HW3 models.

In the holographic gauge $A_z=0$, the tower of pions contributes
to the Chern-Simons action through $A_\mu^\parallel=\partial_\mu \phi$,
whereas in the unitary gauge it appears in $A_z^\mathrm{U}=\partial_z\phi$.
In the latter case, the anomalous interactions of the pions are described by
\be
S^{\text{anom}}_{(\pi)}=\frac{N_{c}}{4 \pi^2}\epsilon^{\mu\nu\rho\sigma}
\,\mathrm{tr}\int d^{4}x\int_{0}^{z_{0}}dz\,A_z^\mathrm{U}\partial_{\mu}V_{\nu}\partial_{\rho}V_{\sigma}-B,
\ee
where $B$ is an infrared subtraction term required by the HW1 model 
and introduced in \cite{Grigoryan:2007wn}, but which disappears in the HW2 and HW3 models
due to their different boundary conditions at $z=z_0$.

With (\ref{AUmodes}), this yields the following pion TFFs (for which $\mathrm{tr}(t^3\mathcal{Q}^2)=1/6$),
\be
F_{\pi_n\gamma^*\gamma^*}(Q_1^2,Q_2^2)=
\frac{N_c}{12\pi^2}g_5 K_n(Q_1^2,Q_2^2),
\ee
with
\be\label{K3n}
K_n(Q_1^2,Q_2^2)=-\int_{0}^{z_{0}}dz\,\mathcal{J}(Q_1,z)\mathcal{J}(Q_2,z)
zy_n(z)+\mathcal{J}(Q_1,z_0)\mathcal{J}(Q_2,z_0)\left(\frac{z}{\beta}\partial_{z}y_n(z)\right)_{z\rightarrow z0},
\ee
where the last term vanishes automatically in the HW3 model.

This can also be written in terms of $\phi$ and $\pi$ modes,
\be
K_n(Q_{1}^{2},Q_{2}^{2})= -\int_{0}^{z_{0}}dz\,\mathcal{J}(Q_{1},z)\mathcal{J}(Q_{2},z)\partial_{z}\phi_n(z)
+\mathcal{J}(Q_{1},z_{0})\mathcal{J}(Q_{2},z_{0})\left[\phi_n(z_0)-\pi_n(z_0)\right] ,
\ee
which implies (equally in the HW1 and HW3 models)
\be
K_n(0,0)=-\pi_n(z_0),
\ee
since $\mathcal{J}(0,z)\equiv 1$.
In the chiral limit one has a constant $\pi_1(z)\equiv -1/(g_5 f_{\pi_1})$ as sole contribution, yielding
$F_{\pi_1 \gamma \gamma}\equiv 
F_{\pi_1 \gamma^* \gamma^*}(0,0) = N_c/(12 \pi^2 f_{\pi_1})$.

We note parenthetically that
the above subtraction term
agrees with the one introduced in \cite{Grigoryan:2007wn}, but the bulk
term therein involves $\partial_z(\phi-\pi)$, which is correct only 
for the $n=1$ mode in the chiral limit
where $\pi_1(z)$ becomes a constant. 
Away from the chiral limit, the expression proposed in
\cite{Grigoryan:2007wn} would in fact give $K_n(0,0)=0$ for all $n$, because
the boundary conditions on the normalizable modes of $\phi$ and $\pi$ are such that they vanish at $z=0$.

\begin{figure}
\includegraphics[width=0.475\textwidth]{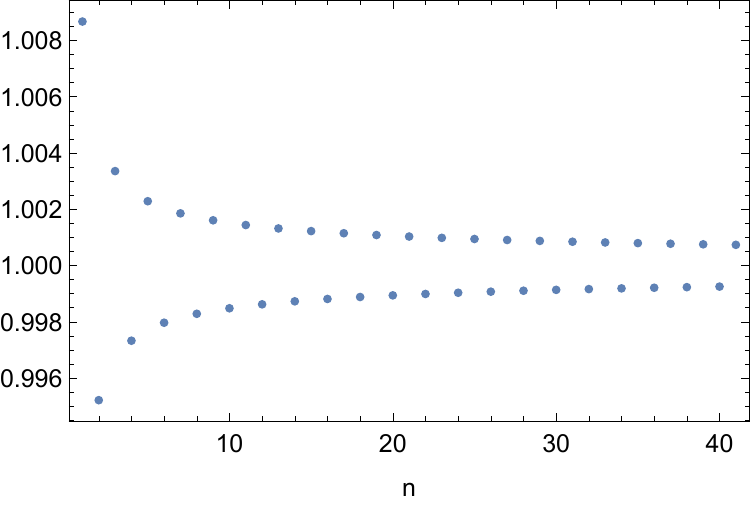}
\caption{Partial sums of the sum relation (\ref{Kn00sum}) as a function of the number of summed modes for the HW1m model.\label{fig:sumr00}}
\end{figure}

With nonzero quark masses, $K_n(0,0)$ can only be given numerically.
However, the sum rule (\ref{sumrule}) implies that $y_S(0,z)=g_5\sum_{n=1}^\infty f_{\pi_n}y_n(z)$
and consequently
\be\label{Kn00sum}
g_5\sum_{n=1}^\infty f_{\pi_n} K_n(0,0)=1,
\ee
or
\be
\sum_{n=1}^\infty f_{\pi_n} F_{\pi_n\gamma\gamma}=\frac{N_c}{12\pi^2},
\ee
showing that then all massive pions contribute to the anomaly.
In Fig.~\ref{fig:sumr00} this is illustrated with the HW1m model (specified
in Sec.~\ref{sec:results}).

However, also in the chiral limit, where $f_{\pi_n}\to0$ for $n>1$ so that the massive pions decouple from the axial
vector current and the axial anomaly, $K_n(0,0)$ and $F_{\pi_n\gamma\gamma}$ remain nonzero.

\section{Short distance constraints}\label{sec:SDC}

\subsection{Form factors}

Remarkably, the short distance constraints of pQCD on meson TFFs \cite{Lepage:1979zb,*Lepage:1980fj,*Brodsky:1981rp,Efremov:1979qk,Hoferichter:2020lap}
including their dependence on the ratio $w=(Q_1^2-Q_2^2)/(Q_1^2+Q_2^2)$
are reproduced exactly by the HW models in the limit $Q_i^2\to\infty$.
With $Q^2=\frac12 (Q_1^2+Q_2^2)\to\infty$ one finds
\be\label{Fpiggasymptotics}
F_{\pi_n\gamma^*\gamma^*}(Q_1^2,Q_2^2)\to \frac{g_5^2 N_c}{12\pi^2} \frac{2f_{\pi_n}}{Q^2}
\left[\frac1{w^2}-\frac{1-w^2}{2w^3}\ln\frac{1+w}{1-w}\right],
\ee
generalizing the chiral result \cite{Grigoryan:2008up} to all massive pions.
Thus with $g_5$ fixed by (\ref{PiVas}), all satisfy the Brodsky-Lepage
limit \cite{Lepage:1979zb,*Lepage:1980fj,*Brodsky:1981rp} with their
respective decay constants,
$\lim_{Q^2\to\infty}Q^2F_{\pi_n\gamma^*\gamma^*}(Q^2,0)=2 f_{\pi_n}$,
and likewise the symmetric pQCD limit \cite{Novikov:1983jt}
$\lim_{Q^2\to\infty}Q^2F_{\pi_n\gamma^*\gamma^*}(Q^2,Q^2)=2 f_{\pi_n}/3$.

The amplitude for axial vector mesons $a_\mu^{(n)}$ decaying into two virtual photons following
from the Chern-Simons action has the form \cite{Leutgeb:2019gbz,Cappiello:2019hwh}
\be\label{calMa}
\mathcal{M}^a=i\frac{N_c}{4\pi^2}\mathrm{tr}(\mathcal{Q}^2 t^a)\,\epsilon_{(1)}^\mu \epsilon_{(2)}^\nu
\epsilon_{A}^{*\rho} \epsilon_{\mu\nu\rho\sigma}
\left[q_{(2)}^\sigma Q_1^2 A_n(Q_1^2,Q_2^2)-q_{(1)}^\sigma Q_2^2 A_n(Q_2^2,Q_1^2)\right],
\ee
where
\begin{align}\label{TFFAn} 
A_n(Q_1^2,Q_2^2) &= \frac{2g_5}{Q_1^2} \int_0^{z_0} dz \left[ \frac{d}{dz} \mathcal{J}(Q_1,z) \right]
\mathcal{J}(Q_2,z) \psi^A_n(z),
\end{align}
which has a finite limit when $Q_1^2\to 0$. Hence,
(\ref{calMa}) vanishes when both photons are real, in accordance
with the Landau-Yang theorem \cite{Landau:1948kw,Yang:1950rg}.

The asymptotic behavior of (\ref{TFFAn}) reads \cite{Leutgeb:2019gbz}
\be
A_n(Q_1^2,Q_2^2) \to 
\frac{g_5^2 F^A_{n}}{Q^4}
\frac1{w^4}\left[
w(3-2w)+\frac12 (w+3)(1-w)\ln\frac{1-w}{1+w}
\right],
\ee
which also agrees with the pQCD behavior derived recently in Ref.~\cite{Hoferichter:2020lap}.

Compared to the most general expression possible for the axial vector
amplitude $\mathcal{M}(a\to\gamma^*\gamma^*)$ \cite{Pascalutsa:2012pr,Roig:2019reh,Zanke:2021wiq}, the holographic result
(\ref{calMa}) has one asymmetric structure function (denoted $C(Q_1^2,Q_2^2)$
in \cite{Roig:2019reh}) set to zero; see Ref.~\cite{Zanke:2021wiq} for
a compilation of the available phenomenological information.

\subsection{Longitudinal short distance constraint on HLBL amplitude}

\subsubsection{$Q_1^2=Q_2^2\gg Q_3^2 \gg m_\rho^2$}

In the Bardeen-Tung-Tarrach basis of the HLBL four-point function \cite{Colangelo:2015ama}, the longitudinal short-distance constraint of Melnikov and Vainshtein \cite{Melnikov:2003xd} in the region
$Q_1^2 \sim Q_2^2\gg Q_3^2 \gg m_\rho^2$ and $Q_4=0$, which is
governed by the chiral anomaly and protected by its nonrenormalization
theorem,
reads \cite{Colangelo:2019lpu,Colangelo:2019uex}
\be\label{MVSDC}
\lim_{Q_3\to\infty} \lim_{Q\to\infty} Q^2 Q_3^2 \bar\Pi_1(Q,Q,Q_3)=-\frac{2}{3\pi^2}
\ee
for $N_c=N_f=3$. 

The short distance behavior of the form factors of both pseudoscalars
and axial vector mesons implies that each individual meson gives a pole contribution
with $\bar\Pi_1(Q,Q,Q_3)\sim Q^{-2} Q_3^{-4}$.
However, in Ref.~\cite{Leutgeb:2019gbz,Cappiello:2019hwh} it was shown that in holographic QCD a summation over the
infinite tower of axial vector mesons changes this. The infinite sum yields
\be
    \bar{\Pi}_1^\mathrm{AV}=-\frac{g_5^2}{2\pi^4 Q_3^2} \int_0^{z_0} dz \int_0^{z_0} dz' \mathcal{J}'(Q,z) \mathcal{J}(Q,z) \mathcal{J}'(Q_3,z') G_A(0;z,z'), \label{HWPi1}
\ee
where $G_A$ is the Green function for the axial vector mode equation
satisfying
\be
\left(z\partial_z\frac1z \partial_z-\beta(z)\right)G_A(0;z,z')=-z\delta(z-z')
\ee
at $q^2=0$.
For large $Q,  Q_3\gg m_\rho$, (\ref{HWPi1}) is dominated by $z, z'\ll z_0$, where one can
approximate $\mathcal{J}(Q,z)\to Qz K_1(Qz)$, and
\be\label{GAsmallz}
G_A(0,z,z') = \frac12 \left(\mathrm{min}(z,z')\right)^2\left(1+\mathcal{O}(Q^{-n})+
\mathcal{O}(Q_3^{-n})\right),\quad n>0,
\ee
when $z=\xi/Q$ and $z'=\xi'/Q_3$.
This asymptotic behavior of $G_A$ holds true in all HW models,
including those with finite quark mass term, because at small $z$ one
has $\beta(z)\sim z^{2(\Delta_- - 1)}$ with $\Delta_-=1$ for the standard choice $M_X^2=-3$,
leading to $n=2$ in (\ref{GAsmallz});
with generalized $M_X^2$, one has $n>0$ as long as $\Delta_- > 0$, which corresponds to $M_X^2<0$.
In all cases, we thus obtain for $Q\gg Q_3\gg m_\rho$
\bea\label{Pi1AVlimit}
&&\lim_{Q_3\to\infty} \lim_{Q\to\infty} Q^2 Q_3^2 \bar\Pi_1^\mathrm{AV}(Q,Q,Q_3) \nonumber\\
&&=-\frac{g_5^2}{2\pi^4}\int_0^\infty d\xi \int_0^\infty d\xi'
\xi K_1(\xi)\frac{d}{d\xi}[\xi K_1(\xi)]\frac{d}{d\xi'}[\xi' K_1'(\xi')]\frac12 \xi^2\nonumber\\
&&=\frac{g_5^2}{(2\pi)^2}\frac1{2\pi^2} \int_0^\infty d\xi  
\xi^2\frac{d}{d\xi}[\xi K_1(\xi)]^2=
-\frac{2}{3\pi^2}
\eea
for $g_5^2=12\pi^2/N_c=(2\pi)^2$, as required by (\ref{MVSDC}).

This already implies that in the massive case the infinite tower of pions (and
the other pseudo-Goldstone bosons) should not contribute to the Melnikov-Vainshtein constraint, i.e., summing the infinite number of contributions should not change
the asymptotic behavior of the individual contributions in the same way as it
happens with axial vector mesons.
In order to check this, we need to analyse
\bea
\bar\Pi_1^{(\pi)}&=&-\sum_{n=1}^\infty \frac{F_{\pi_n\gamma^*\gamma^*}(Q_1^2,Q_2^2)F_{\pi_n\gamma^*\gamma}(Q_3^2)}{Q_3^2+m_n^2}\nonumber\\
&=&-\frac{1}{4\pi^2}\sum_n \frac1{Q_3^2+m_n^2}\biggl\{\mathcal{J}(Q_{1},z_{0})\mathcal{J}(Q_{2},z_{0})\mathcal{J}(Q_{3},z_{0})\left(\frac{z}{\beta}\partial_{z}y_{n}(z)\right)_{z\rightarrow z0}^2\nonumber
\\\nonumber
&&-\mathcal{J}(Q_{1},z_{0})\mathcal{J}(Q_{2},z_{0})\left(\frac{z}{\beta}\partial_{z}y_{n}(z)\right)_{z\rightarrow z0}\int_0^{z_0} dz\mathcal{J}(Q_{3},z) zy_{n}(z)\\\nonumber
&&-\mathcal{J}(Q_{3},z_{0})\left(\frac{z}{\beta}\partial_{z}y_{n}(z)\right)_{z\rightarrow z0}\int_0^{z_0} dz\mathcal{J}(Q_{1},z)\mathcal{J}(Q_{2},z)zy_{n}(z)\\
&& +\int_0^{z_0} dz\mathcal{J}(Q_{1},z)\mathcal{J}(Q_{2},z)zy_{n}(z) \int_0^{z_0} dz'\mathcal{J}(Q_{3},z') z' y_{n}(z')
\biggr\}.
\eea
Since $\lim_{Q\rightarrow\infty} \mathcal{J}(Q,z_{0})=0$ and $\lim_{Q\rightarrow\infty} Q^2\mathcal{J}(Q,z_{0})=0$, the first three terms in the curly brackets do not survive in
the large-$Q$ limit. The last term can be formally written as
\bea
&&-\frac{1}{4\pi^2}\sum_{n=1}^\infty\int_0^{z_0} dz dz' \mathcal{J}(Q_{1},z)\mathcal{J}(Q_{2},z)\mathcal{J}(Q_{3},z')zz' y_{n}(z')
(Q_3^2-L)^{-1}y_{n}(z)  \nonumber \\ 
&=& -\frac{1}{4\pi^2}\int_0^{z_0} dz dz'\mathcal{J}(Q_{1},z)\mathcal{J}(Q_{2},z)  \mathcal{J}(Q_{3},z') z' z (Q_3^2-L)^{-1} G(z,z';0),
\label{eq:Pi1piasympt}
\eea
where $L$ is the differential operator introduced in (\ref{eq:eomy}) and $G$ its
Green function as defined in (\ref{Greendef}). In the limit of interest for the Melnikov-Vainshtein
constraint, one has $z'\gg z$, thus (\ref{Greendef}) reduces to $(L+q^2)G(z,z';q)=0$; hence,
effectively $LG(z,z';0)=0$ and we obtain
\begin{align}
\bar{\Pi}_1^{(\pi)}\Big|_{Q \gg Q_3\gg m_\rho} \to -\frac{1}{4\pi^2 Q_3^2}\int_0^{z_0} dz dz' \mathcal{J}(Q_{1},z)\mathcal{J}(Q_{2},z)  \mathcal{J}(Q_{3},z') z' z G(z,z';0).
\end{align}
At parametrically small $z=\xi/Q$ and $z'=\xi'/Q_3$ and at $q^2=0$, (\ref{Greendef}) reduces to
\be
\partial_z \frac{z}{\beta(z)}\partial_z G(0;z,z')=-\delta(z-z'),
\quad \beta(z)\sim M_q^2(z/z_0)^{2-2\Delta_-},
\ee
provided $\Delta_- > 0$.

In the massive HW models with $M_X^2=-3$, this gives
\be
G(0;z,z') \to -{M_q^2} \ln\left(\mathrm{max}(z,z')\right)+\mathrm{const.},
\ee
leading to
\be\label{Pi1piasymptotic}
-Q^2 Q_3^2 \bar\Pi_1^{(\pi)}(Q,Q,Q_3) 
\sim\frac{M_q^2}{4\pi^2}\frac{\ln(Q_3)}{Q_3^2}
\int_0^\infty d\xi \xi^3 [K_1(\xi)]^2 
\int_0^\infty d\xi'\xi'^2 K_1(\xi')
=\frac{M_q^2}{6\pi^2}\frac{\ln(Q_3^2)}{Q_3^2}
\to 0
\ee
for large $Q\gg Q_3\gg m_\rho$.
Thus the summation of the infinite tower of pions does give a different asymptotic
behavior than that of individual contributions, but only in the form of a logarithmic enhancement.

\begin{figure}
\includegraphics[width=0.48\textwidth]{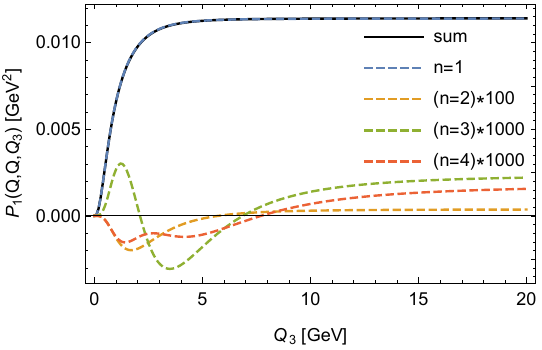}\qquad
\includegraphics[width=0.473\textwidth]{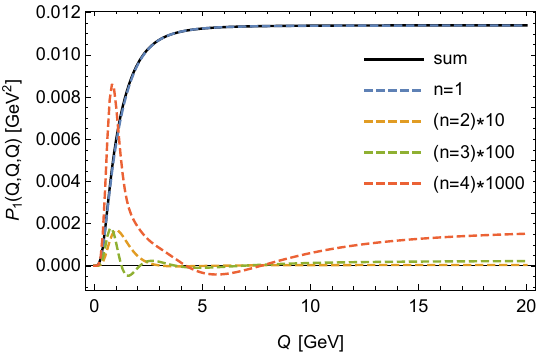}
\caption{Plot of the contribution of the first four pion modes to $P_1(Q,Q,Q_3)=-Q^2 Q_3^4 \Pi_1^{(\pi)}(Q,Q,Q_3)$ for the HW1m model with $Q=200$ GeV in the left panel and $Q_3=Q$ in the right panel, showing that the contribution from the tower of pseudoscalars to the LSDC is suppressed like $1/Q_3^2$ up to (here invisible) logarithmic corrections.}
\label{fig:LSDC}
\end{figure}

\begin{figure}
\includegraphics[width=0.475\textwidth]{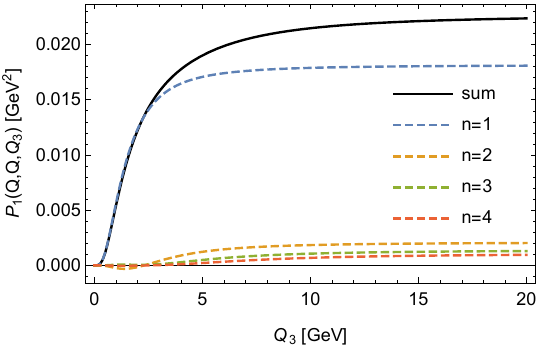}\qquad
\includegraphics[width=0.475\textwidth]{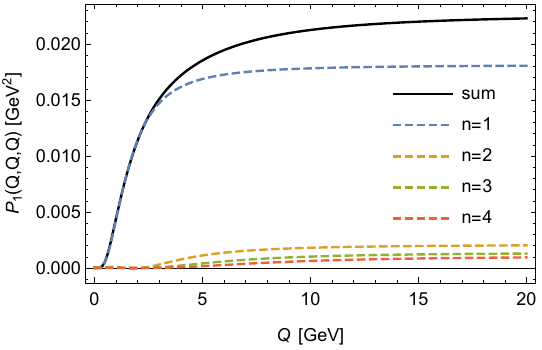}
\caption{As in Fig.~\ref{fig:LSDC} but with quark masses increased by a factor of 25, illustrating the the build-up of a logarithmic enhancement by the summation of the infinite tower of massive pions.}
\label{fig:LSDCmqx}
\end{figure}

In Fig.~\ref{fig:LSDC}, the contributions of the first four pion modes to $P_1\equiv -Q^2 Q_3^4 \bar\Pi_1^{(\pi)}(Q,Q,Q_3)$, i.e., the left hand side of (\ref{Pi1piasymptotic}) with another factor of $Q_3^2$, is plotted for the HW1m model with $Q=200$ GeV and increasing $Q_3$.
Numerically, this is dominated by the contribution from the lightest pion
which completely swamps the logarithmic term (\ref{Pi1piasymptotic}) whose
prefactor $M_q^2/(3\pi^2)$ is of the order $\sim 3\cdot 10^{-6}\,\mathrm{GeV}^2$. Note that this term is
suppressed by an extra power of quark mass compared to the contribution
from the magnetic susceptibility \cite{Vainshtein:2002nv,Gorsky:2009ma}
to the asymptotic behavior of $\bar\Pi_1$ worked out in Ref.~\cite{Bijnens:2019ghy}.
In Fig.~\ref{fig:LSDCmqx} the same is plotted with quark masses increased by a factor of 25 (corresponding to a pion mass of about 750 MeV), where the slow build-up of a logarithmic term becomes apparent.

When $M_X^2<-3$, the logarithmic enhancement disappears
and $Q^2 Q_3^2 \bar\Pi_1^{(\pi)}(Q,Q,Q_3)\sim Q_3^{-2}$.
For $-3<M_X^2<0$, where $1>\Delta_->0$, one instead obtains a power-law enhancement
from
\be\label{G0asymptotics}
G(0;z,z')\to \frac{M_q^2}{2(1-\Delta_-)}\left( \frac{z_0}{z'} \right)^{2(1-\Delta_-)} \propto Q_3^{2(1-\Delta_-)}
\ee
such that 
\be
Q^2 Q_3^2 \bar\Pi_1^{(\pi)}(Q,Q,Q_3) \sim -\frac{2^{2\Delta^- - 3}\Gamma(\Delta^-)\Gamma(1+\Delta^-)}{3\pi^2}
\frac{M_q^2}{Q_3^2}
\frac{(z_0 Q_3)^{2(1-\Delta^-)}}{1-\Delta^-} \propto Q_3^{-2\Delta^-}.
\ee
Only for $\Delta_-=0$, which is at the border of the allowed
range $M_X^2\in(-4,0)$, would the infinite tower of massive
pions start to contribute to the Melnikov-Vainshtein constraint,
exactly when the result (\ref{Pi1AVlimit}) for the infinite tower of axial vector mesons
would break down. However, in the following applications
the phenomenologically interesting generalizations of $M_X^2$
all have $M_X^2<-3$, where the contribution from the pseudoscalar tower to
the longitudinal short distance constraint is suppressed by two inverse powers of $Q_3$ without even a logarithmic enhancement.

In the chiral limit, the massive pions still contribute to $\bar\Pi_1$, even though they decouple from the axial vector current and from the axial anomaly. With strictly $M_q=0$,
\be
G(0;z,z')_\mathrm{chiral}\to \mathrm{const.}
+ \frac{\Sigma^2 z_0^4}{2(1-\Delta_+)}\left( \frac{z_0}{z'} \right)^{2(1-\Delta_+)} \propto \mathrm{const.} + Q_3^{-2-2\alpha},
\ee
so in this case there is no enhancement from the summation of the infinite tower, irrespective of the value of $M_X^2$.

\subsubsection{$Q_1^2=Q_2^2=Q_3^2 \gg m_\rho^2$}

In the symmetric limit $Q_1^2=Q_2^2=Q_3^2 \gg m_\rho^2$, operator product expansion and LO pQCD imply that the
longitudinal short distance constraint is 2/3 of value appearing in (\ref{MVSDC}) \cite{Melnikov:2003xd,Bijnens:2019ghy},
\be\label{symSDC}
\lim_{Q\to\infty} \lim_{Q\to\infty} Q^4 \bar\Pi_1(Q,Q,Q)=-\frac{4}{9\pi^2}.
\ee
In this case, the same derivation as above leads to
\bea\label{Pi1AVsymlimit}
&&\lim_{Q\to\infty} Q^4 \bar\Pi_1^\mathrm{AV}(Q,Q,Q) \nonumber\\
&&=-\frac{g_5^2}{2\pi^4}\int_0^\infty d\xi \int_0^\infty d\xi'
\frac12\frac{d}{d\xi}[\xi K_1(\xi)]^2\frac{d}{d\xi'}[\xi' K_1'(\xi')]\frac12 [\mathrm{min}(\xi,\xi')]^2\nonumber\\
&&=-\frac{g_5^2}{(2\pi)^2 \pi^2}\int_0^\infty d\xi \int_0^\infty d\xi'
[\xi K_1(\xi)]^2[\xi' K_1'(\xi')]\xi\delta(\xi-\xi')\nonumber\\
&&=-\frac{g_5^2}{(2\pi)^2\pi^2} \int_0^\infty d\xi  
\,\xi [\xi K_1(\xi)]^3=
-0.361\frac{g_5^2}{(2\pi)^2\pi^2},
\eea
which reproduces the correct power behavior, but for $g_5=2\pi$,
as demanded by (\ref{PiVas}), the numerical value is only 81\% of the result (\ref{symSDC}). Again, this result for the contribution of the infinite tower
of axial vector mesons is the same for chiral and for massive HW models.

The asymptotic contribution of the pseudoscalar tower is still
given by the expression (\ref{eq:Pi1piasympt}), but the argument given thereafter
does not apply in the symmetric limit.
However, numerically we found no evidence of an enhancement
of the asymptotic behavior due to the summation of the infinite tower
of massive pions beyond what is seen in the asymmetric case, see Figs.~\ref{fig:LSDC} and \ref{fig:LSDCmqx}.

\section{Numerical Results}\label{sec:results}

In the following we compare the different HW models numerically,
in particular with regard to the contribution of the infinite tower of
massive pions to the hadronic light-by-light scattering amplitude
and thereby to the anomalous magnetic moment of the muon.

In all models we have chosen $g_5=2\pi$, ensuring an exact fit of the asymptotic LO pQCD
result (\ref{PiVas}), but in the later discussion we shall also consider relaxing this
constraint to account for the fact that at any large but finite energy scale
there is a nonnegligible reduction of TFFs
of the order $\alpha_s/\pi$ that in AdS/QCD could perhaps be simply\footnote{More elaborate holographic QCD models would do this by a modification
of the anti-de Sitter background.}
modelled by a small reduction of $g_5$. The low-energy parameters $f_\pi$ and
$m_\rho$ are always fitted to their physical values, fixed to 92.4 MeV and 775 MeV, respectively, to be consistent with our previous work \cite{Leutgeb:2019gbz}.


We consider two possibilities (HW1, HW3) for boundary conditions at $z=z_0$ [always given
by (\ref{eq:z0})], and also alternatively the standard choice $M_X^2=-3$ and
having $M_X^2$ as a tunable parameter. We thus employ four different HW models with nonvanishing light
quark masses.

\begin{description}
    \item[HW1m] is the direct extension of the chiral HW1 model employed in our previous
    studies \cite{Leutgeb:2019zpq,Leutgeb:2019gbz}. It coincides with
    model A of Erlich, Katz, Son and Stephanov \cite{Erlich:2005qh} except that we have fitted to the mass of the neutral instead of the charged pions;
    \item[HW1m'] deviates from the standard choice $M_X^2=-3$ in order to attempt a better fit of the masses of the first excited pion and/or the lightest axial vector meson. It turns out that only the latter can be matched to the $a_1$ mass. The mass of the first excited pion is then also reduced compared to the pristine HW1 model, from around 1900 to 1600 MeV, but the target of 1300 MeV cannot be reached.
    \item[HW3m] uses the standard choice $M_X^2=-3$, but boundary conditions as proposed in \cite{Domenech:2010aq}, which have the advantage of making a manual subtraction of infrared boundary contributions in the Chern-Simons action unnecessary;
    \item[HW3m'] uses additionally $M_X^2$ as a free parameter, which in this model achieves a fit of the mass of the first excited pion of 1300 MeV, as was the main motivation put forward by Dom\`enech, Panico, and Wulzer in Ref.~\cite{Domenech:2010aq} for proposing this kind of model. Our HW3m' slightly deviated from the parameters used in Ref.~\cite{Domenech:2010aq} because we fitted $f_\pi$, $m_\rho$, $m_{\pi^0}$, and $m_{\pi(1300)}$ instead of performing a least-squares fit over a larger set of low-energy parameters.
\end{description}

For the chiral limit, we only consider the HW1 model, where we update
the results obtained for pions in \cite{Leutgeb:2019zpq}\footnote{Note that the original version of \cite{Leutgeb:2019zpq} contained an error in the $a_\mu$ results for the HW1 model, but not in the plots of the corresponding TFFs.} and recapitulate the results for axial vector mesons in \cite{Leutgeb:2019gbz}.

\subsection{Masses}

As Table \ref{tab:hwmodels} shows, in the chiral HW1 model the mass of the lightest axial-vector multiplet is above the physical masses \cite{PDG20} $M_{a_1(1260)}=1230(40)$ MeV and $M_{f_1(1285)}=1281.9(5)$ MeV, 
but below that of the $f_1'$, $M_{f_1(1420)}=1426.3(9)$ MeV. While this prediction of the HW1 model
is in the right ballpark,
the mass of the first excited pion is 1899 MeV
and thus significantly above $M_{\pi(1300)}=1300\pm100$ MeV.

With nonzero light quark masses, the HW1m model has slightly reduced axial-vector meson mass and
increased excited pion masses. In the HW1m', where the $a_1$ mass can be matched, the lowest excited pion mass is reduced to about 1600 MeV.

With HW3 boundary conditions, there is additional chiral symmetry
breaking and therefore a larger difference between the vector and axial vector meson masses, 
so that $m_{a_1}$ is now even above the mass of the physical $f_1'$,
but the excited pion mass is lowered substantially compared to the HW1m model,
while still being too high. 
In the HW3m' model, where the first excited pion can be
brought down to 1300 MeV, the axial vector meson mass is then also lowered, but somewhat larger than
in the HW1 models.
Even in the HW3m' model, where the pseudoscalar masses are the smallest, the second excited ($n=3$)
pion has a mass higher than the next established pion state\footnote{Note, however, that this state is sometimes considered to be a non-$q\bar q$ state \cite{PDG20}.} $\pi(1800)$ and instead close to the next (less established) state $\pi(2070)$; in the other
models $m_{n=3}$ is far higher. 

\subsection{Decay constants}

The results for the decay constant of the lowest excited pion in the massive HW models span the
range (1.56\ldots 1.92) MeV, with the HW3m' model, where the mass of $\pi(1300)$ can be fitted,
yielding the largest value. These values are well below the existing experimental
upper bound of 8.4 MeV \cite{Diehl:2001xe}; 
the result for the HW3m' model is actually 
fully consistent with the value 2.20(46) MeV of Ref.~\cite{Maltman:2001gc} that was adopted in
Ref.~\cite{Colangelo:2019uex}.%
\footnote{The decay constants of the higher pion modes fall off with increasing mode number $n$, inversely proportional to $m_n^x$, where $x$ is smaller than 1, which is very different from the value $n=2$ of \cite{Elias:1997ya} and closer to but still not agreeing with \cite{Kataev:1982xu} where $x=1$. However, the HW models lack linear Regge
trajectories; soft-wall models may be more realistic here.}

The holographic results for the decay constant of the lightest axial vector meson, defined
in analogy to (\ref{FVdef}),
read $F_{a_1}=(493\,\text{MeV})^2$ in the chiral HW1 model and $(426\ldots 506\,\text{MeV})^2$
for the different massive HW models. 
Note that in the literature frequently the mass of the axial vector
meson is factored out \cite{Ecker:1988te,*Ecker:1989yg}.\footnote{Our definition of $F_{a_1}$ corresponds to $F_A^{a=3}m_A$ and $F_A^{a=3}m_A/\sqrt2$ in \cite{Hoferichter:2020lap,Zanke:2021wiq} and \cite{Yang:2007zt}, respectively.}
In Ref.~\cite{Yang:2007zt} a value of $F_{a_1}/m_{a_1}=168(7)$ MeV has been obtained
from light-cone sum rules.
With $F_{a_1}/m_{a_1}=177$ MeV
for the chiral HW1 model and (148\ldots195) MeV for the different massive HW models (see Table \ref{tab:hwmodels}), the ballpark spanned by the holographic results is broadly consistent
with that.

\newcommand{\AV}{\mathrm{A}}
\begin{table}
\begin{center}
\begin{tabular}{  l  c c  c  c  c  c  c  c c c } 
\toprule
model & PS & $n=1$ & $n=2$ & $n=3$ & AV & $n=1$ & $n=2$ & $n=3$& $n=4$ & $n=5$ \\
\hline
\hline
HW1 chiral 
& $m_{\pi_n}${\tiny[MeV]} & $0\Rsh135$ & 1899 & 2887 & $m_{\AV_n}${\tiny[MeV]} & 1375 & 2154 & 2995 & 3939 & 4917 \\ 

\cline{2-11} 
$M_X^2 = -3$ 
 & $f_{\pi_n}$ {\tiny[MeV]} & 92.4* & 0 & 0 & $F_{\AV_n}/m_{\AV_n}$  {\tiny[MeV]}  & 177 & 204 & 263 & 311 & 351\\

\cline{2-11} 
 & $F_{\pi_n\gamma\gamma}${\tiny[GeV$^{-1}$]} & 0.274 & -0.202 & 0.153 & $A(0,0)$ {\tiny[GeV$^{-2}$]}  & -21.04 & -2.93 & 0.294 & -2.16 & 0.400  \\ 
 
\cline{2-11} 
& $a_\mu^{\pi_n}\cdot 10^{11}$ & 65.2 & 0.7 & 0.1 & $a_\mu^{\AV_n}\cdot 10^{11}$ & 31.4 
& 4.7 & 1.8 & 1.2 & 0.5  \\

\hline
\hline

HW1m 
& $m_{\pi_n}${\tiny[MeV]} & 135* & 1892 & 2882 & $m_{\AV_n}${\tiny[MeV]} & 1367 & 2141 & 2987 & 3934 & 4914 \\ 

\cline{2-11} 
$M_X^2 = -3$ 
& $f_{\pi_n}$ {\tiny[MeV]} & 92.4* & 1.56 & 1.25 & $F_{\AV_n}/m_{\AV_n}$  {\tiny[MeV]}  & 175 & 204 & 263 & 311 & 351 \\

\cline{2-11} 
 & $F_{\pi_n\gamma\gamma}${\tiny[GeV$^{-1}$]} & 0.276 & -0.203 & 0.154 & $A(0,0)$ {\tiny[GeV$^{-2}$]}  & -21.00 & -3.21 & 0.328 & -2.16 & 0.376  \\ 

\cline{2-11}

& $a_\mu^{\pi_n}\cdot 10^{11}$ & 66.0 & 0.7 & 0.1 & $a_\mu^{\AV_n}\cdot 10^{11}$ & 31.4  
& 4.9 & 1.8 & 1.2 & 0.5 \\

\hline
\hline

HW1m' 
& $m_{\pi_n}${\tiny[MeV]} & 135* & 1591 & 2564 & $m_{\AV_n}${\tiny[MeV]} & 1230* & 1977 & 2901 & 3879 & 4873\\ 

\cline{2-11} 
$M_X^2 = -3.837$ 
 & $f_{\pi_n}$ {\tiny[MeV]} & 92.4* & 1.59 & 0.950 & $F_{\AV_n}/m_{\AV_n}$  {\tiny[MeV]}  & 148 & 208 & 266 & 312 & 351  \\

\cline{2-11} 
 & $F_{\pi_n\gamma\gamma}${\tiny[GeV$^{-1}$]}  & 0.277 & -0.250 & 0.194 & $A(0,0)$ {\tiny[GeV$^{-2}$]}  & -19.95 & -7.29 & 0.678 & -2.18 & 0.341 \\ 
 
\cline{2-11} 
& $a_\mu^{\pi_n}\cdot 10^{11}$ & 64.3 & 1.5 & 0.3 & $a_\mu^{\AV_n}\cdot 10^{11}$ & 29.8 
& 8.7 & 2.0 & 1.3 & 0.5 \\

\hline
\hline

HW3m & $m_{\pi_n}${\tiny[MeV]} 
& 135* & 1715 & 2513 & $m_{\AV_n}${\tiny[MeV]} & 1431 & 2421 & 3398 & 4387 & 5384 \\ 

\cline{2-11} 
$M_X^2 = -3$ 
 & $f_{\pi_n}$ {\tiny[MeV]} & 92.4* & 1.56 & 1.34 & $F_{\AV_n}/m_{\AV_n}$  {\tiny[MeV]}  & 195 & 244 & 291 & 332 & 369 \\

\cline{2-11} 
 & $F_{\pi_n\gamma\gamma}${\tiny[GeV$^{-1}$]}  & 0.277 & -0.196 & 0.0797 & $A(0,0)$ {\tiny[GeV$^{-2}$]}  & -21.27 & 0.310 & -2.09 & -0.299 & -0.514 \\ 
 
\cline{2-11} 

& $a_\mu^{\pi_n}\cdot 10^{11}$ & 66.6 & 0.8 & 0.04 & $a_\mu^{\AV_n}\cdot 10^{11}$ & 32.7 
& 3.4 & 1.7 & 0.7 & 0.4 \\

\hline
\hline

HW3m' 
& $m_{\pi_n}${\tiny[MeV]} & 135* & 1300* & 2113 & $m_{\AV_n}${\tiny[MeV]} & 1380 & 2355 & 3345 & 4345 & 5350\\ 
\cline{2-11} 
$M_X^2 = -3.841$ 
 & $f_{\pi_n}$ {\tiny[MeV]} & 92.4* & 1.92 & 1.29 & $F_{\AV_n}/m_{\AV_n}$  {\tiny[MeV]}  & 186 & 242 & 291 & 332 & 369 \\

\cline{2-11} 
 & $F_{\pi_n\gamma\gamma}${\tiny[GeV$^{-1}$]}  & 0.278 & -0.206 & 0.0474 & $A(0,0)$ {\tiny[GeV$^{-2}$]}  & -21.29 & -0.841 & -1.76 & -0.440 & -0.476 \\ 
 
\cline{2-11} 

& $a_\mu^{\pi_n}\cdot 10^{11}$ & 66.0 & 1.5 & 0.01 & $a_\mu^{\AV_n}\cdot 10^{11}$ & 33.2 
& 4.1 & 1.8 & 0.8 & 0.4 \\

\botrule
\end{tabular}
\end{center}
\caption{Numerical results for various HW models (in the chiral HW1 model with the pion mass raised manually ($\Rsh$) to 135 MeV in the evaluation of $a_\mu$). Fitted values are marked by $*$. The axial vector contributions correspond to a whole $U(N_f=3)$ multiplet, which in the present flavor-symmetric case is simply
$a_\mu^{A_n}=a_\mu^{(a_1+f_1+f_1')_n}=4 a_\mu^{{a_1}_n}$.}
\label{tab:hwmodels}
\end{table}

\subsection{Comparison of transition form factors}\label{sec:compareTFF}

\begin{figure}
\includegraphics[width=0.475\textwidth]{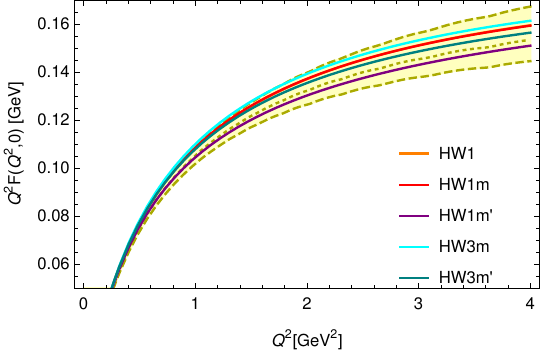}\qquad
\includegraphics[width=0.475\textwidth]{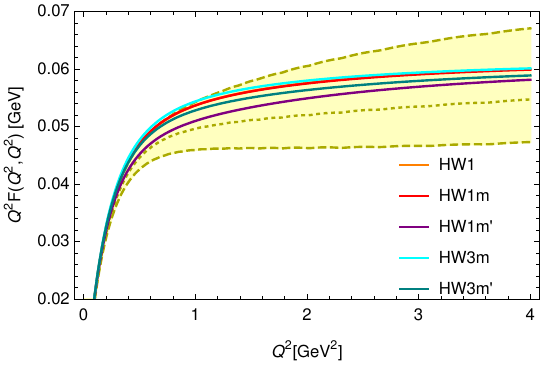}
\caption{Single and double-virtual pion transition form factors in the various HW models together with the 
result of the data-driven dispersive approach of Ref.~\cite{Hoferichter:2018kwz}
(yellow band for the estimated error with dotted dark-yellow points for the central result). 
}
\label{fig:PionTFF}
\end{figure}

\begin{figure}
\includegraphics[width=0.475\textwidth]{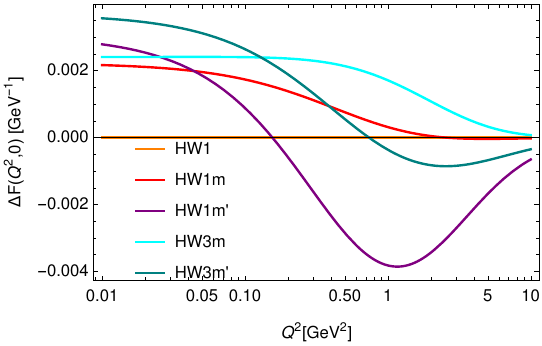}\qquad
\includegraphics[width=0.475\textwidth]{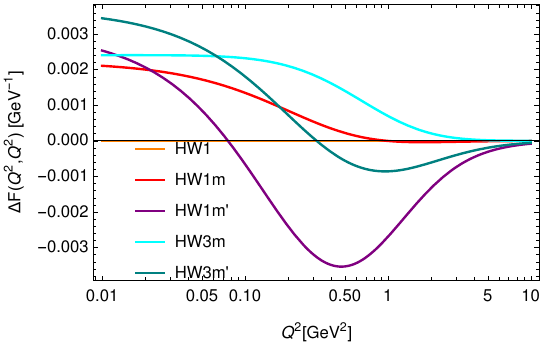}
\caption{
\label{fig:DPionTFF}Deviations of the single and double-virtual pion transition form factors in the massive HW models from the chiral HW1 model.}
\end{figure}

In Figs.~\ref{fig:PionTFF} and \ref{fig:DPionTFF} the single and double-virtual pion TFF following from the chiral and the massive HW models
are compared with each other and with the results obtained in \cite{Hoferichter:2018kwz} in the dispersive approach \cite{Colangelo:2015ama}. The HW results all lie within the error band of the dispersive result, mostly above its central value, with only HW1m' below.

Table \ref{tab:hwmodels} also lists the amplitude $F_{\pi_n\gamma\gamma}$ for pseudoscalar decays into two real photons. For the ground state pion there is only a tiny change when finite quark masses are introduced, across all massive HW models. For excited pions there is more variation; for the first
excited pion the range of $|F_{\pi_2\gamma\gamma}|$ is (0.196\ldots0.250) GeV$^{-1}$. In the HW3m' model, where one can fit to $m_2=1300$ MeV, the value is 0.206 GeV$^{-1}$.
At present, no direct measurements are available, but data on certain branching ratios
permit to derive an estimate for an upper bound \cite{Colangelo:2019uex} reading
$|F_{\pi(1300)\gamma\gamma}|<0.0544(71) \text{Gev}^{-1}$. Evidently, the
holographic results strongly overestimate the two-photon coupling of the first excited pion. This could be taken as a hint that it is better to choose a model where the masses
of the lightest axial vector mesons can be fitted to experiment, namely the HW1m' model,
since the two-photon couplings of axial vector mesons play a more important role
altogether. The holographic results obtained for the whole tower of excited pions could still be taken as a rough estimate of (perhaps an upper bound of) this contribution in real QCD. It is certainly also conceivable that the
contributions of individual modes are overestimated while their combined
contribution is closer to real QCD, where the mass spectrum of
excited pions is denser than in the HW models.

\begin{figure}
\includegraphics[width=0.475\textwidth]{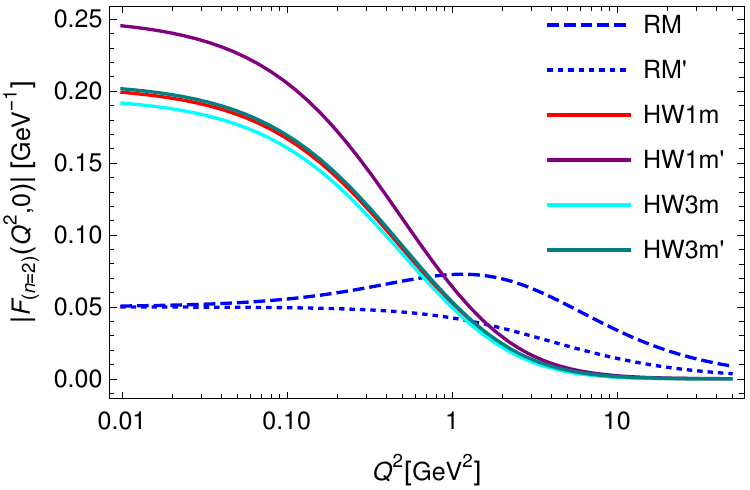}\qquad
\includegraphics[width=0.475\textwidth]{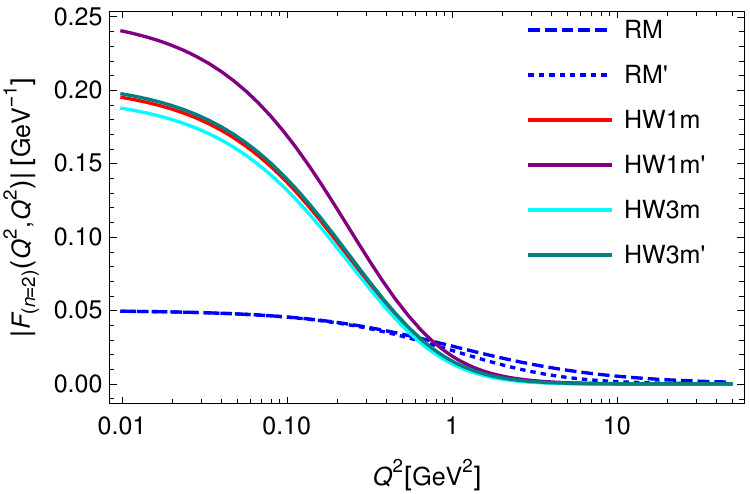}
\caption{
\label{fig:Pion1300TFF}
Single and double-virtual TFFs for the first excited pion in the various HW models. Also given are the corresponding quantities in the Regge model of Ref.~\cite{Colangelo:2019lpu,Colangelo:2019uex} (RM) and its modification according to App.\ E therein (RM'). 
}
\end{figure}

\begin{figure}
\includegraphics[width=0.475\textwidth]{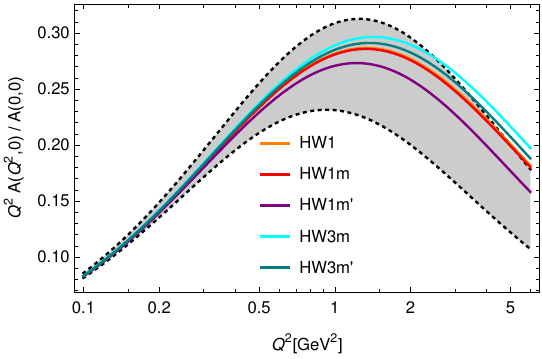}\qquad
\includegraphics[width=0.475\textwidth]{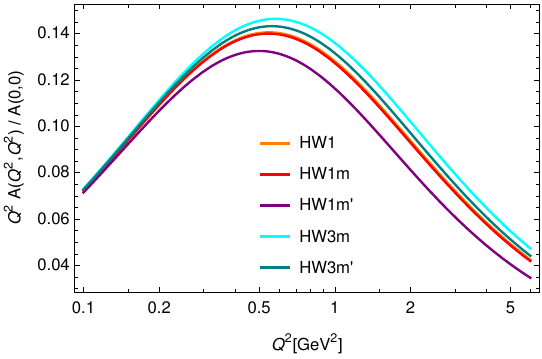}
\caption{Single and double-virtual axial vector transition form factors, the former compared with the dipole fit of L3 data for $f_1(1285)$ of Ref.~\cite{Achard:2001uu} (grey band).}
\label{fig:AVTFF}
\end{figure}

Table \ref{tab:hwmodels} also shows that the chiral HW1 model has almost identical results for $F_{\pi_n\gamma\gamma}$ as the HW1m model. As
emphasized in Ref.~\cite{Colangelo:2019uex}, the vanishing of the
decay constants of massive pions in the chiral limit does not
preclude a coupling to photons. The former only means decoupling of the massive
pions from
the axial vector current and the axial anomaly. Thus massive pions contribute
to the HLBL scattering amplitude also in the chiral HW1 model and therefore to the anomalous magnetic moment
of the muon.\footnote{The simpler HW2 model considered in Refs.~\cite{Leutgeb:2019gbz,Cappiello:2019hwh} does not involve a bifundamental bulk scalar field and thus does not have excited pion states at all.}

Fig.~\ref{fig:Pion1300TFF} displays the single and double-virtual TFFs for the first excited pion in the various HW models. Also shown are the corresponding quantities in the Regge model of Ref.~\cite{Colangelo:2019lpu,Colangelo:2019uex} (RM) and its modification according to App.\ E therein (RM').
The Regge model TFFs satisfy the experimental upper bound $F_{\pi(1300)\gamma\gamma}$, but
do not obey the single-virtual Brodsky-Lepage limit with the decay constants
of excited pseudoscalars. The HW results have a much larger amplitude at
vanishing virtualities, but decay faster. They have a zero and a sign change before approaching the asymptotic behavior (\ref{Fpiggasymptotics}). The TFFs of higher excited pions have more zeros and reduced asymptotic limits due to smaller
decay constants.

In Fig.~\ref{fig:AVTFF} the shape of the axial vector TFF $A(Q_1^2,Q_2^2)/A(0,0)$ is shown
in the single-virtual and in the symmetric double-virtual cases.
In the single-virtual case we also show the experimental result obtained
in \cite{Achard:2001uu} for the $f_1(1285)$ meson which is found to be remarkably
compatible, in particular for the HW1m' model, where the mass of the
lightest axial vector meson can be fitted.

The normalization factor $A(0,0)$ is directly related to the so-called equivalent
two-photon rate \cite{Schuler:1997yw,Pascalutsa:2012pr}. In Ref.~\cite{Leutgeb:2019gbz} we have compared with a combination
of the experimental results obtained for the $f_1(1285)$ and $f_1(1420)$ meson.
Since we are actually using flavor-symmetric models, one should account for
SU(3) flavor breaking before matching $A(0,0)$,\footnote{We are grateful to
Martin Hoferichter for pointing this out to us.}
which leads to
$|A(0,0)|_{f_1(1285)}^\mathrm{exp.}=16.6(1.5)\,\text{GeV}^{-2}$.
In Refs.~\cite{Roig:2019reh,Masjuan:2020jsf} the corresponding value for the lightest $a_1$ meson
has been estimated as
$|A(0,0)|^\mathrm{exp}_{a_1(1260)}=19.3(5.0)\,\text{GeV}^{-2}$. In the HW models we find the
range $|A(0,0)|_{n=1}=(19.95\ldots21.29)\,\text{GeV}^{-2}$ which is compatible
with the latter, but above the value obtained for the $f_1(1285)$ meson.

\subsection{HLBL contribution to $a_\mu$}

\begin{figure}[b]
\includegraphics[width=0.55\textwidth]{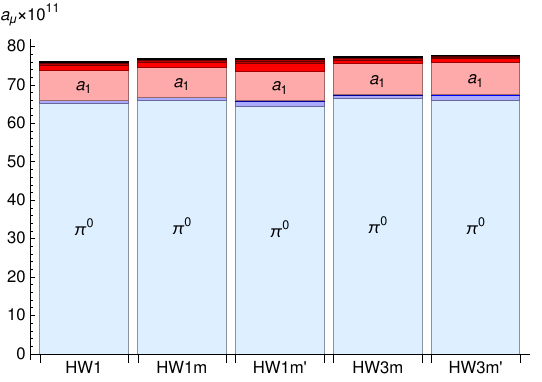}
\caption{Bar chart of the individual contributions to $a_\mu^{\pi\cup a_1}$
in the various HW models, with excited modes given by increasingly darker colors, blue for the $\pi^0$'s, red for the $a_1$'s.}
\label{fig:barchart}
\end{figure}

In Table \ref{tab:hwmodels} we also give the holographic results for
the contributions to $a_\mu$, the anomalous magnetic moment
of the muon, from the first few states of the pion and axial vector meson towers;
in Fig.~\ref{fig:barchart} the results for the $\pi^0$ and $a_1$ sector
are shown in form of a bar chart.

In the chiral HW1 model, the chiral TFFs for the pion have been combined
with a pion propagator where the physical mass of $\pi^0$ has been inserted
by hand. The result of $a_\mu^{\pi^0}=65.2\times 10^{-11}$ is remarkably
close to that obtained in the massive HW models, which together span the
range $a_\mu^{\pi^0}=(64.3\ldots66.6)\times 10^{-11}$. The results of the massive HW1m
are very close to those of the chiral HW1 model, also for the other contributions.
Somewhat more variation is obtained in the other models, where either different
boundary conditions or different values of $M_X^2$ are employed.

\begin{table}
\bigskip
\begin{tabular}{lcccccccc}
\toprule
model\phantom{$^{1^{1^1}}$} & $a_\mu^{\pi^0}$ & $a_\mu^{\pi}$ & $a_\mu^{\pi\cup a_1}$ & $a_{\mu(L)}^{\pi^*\cup a_1}$ & $a_{\mu(L)}^{\pi^*\cup a_1^*}$ & $a_\mu^{\mathrm{P}^*}$ & $a_\mu^\mathrm{A}[L+T]$ & $a_\mu^{\mathrm{A}+\mathrm{P}^*}$
\\
\colrule
HW1 & 65.2 & 66.1 & 76.2 & 6.6 & 2.2 
& 3.5 & 40.6[23.2+17.4] & 44.0 \\
HW1m & 66.0 & 66.8 & 77.0 & 6.7 & 2.2
& 3.5 & 40.8[23.3+17.5] & 44.3 \\
HW1m' & 64.3 & 66.1 & 77.0 & 8.1 & 3.7
& 7.2 & 43.3[25.0+18.3] & 50.5 \\
HW3m & 66.6 & 67.4 & 77.4 & 6.5 & 1.9
& 3.4 & 39.9[22.7+17.2] & 43.3 \\
HW3m' & 66.0 & 67.5 & 77.8 & 7.4 & 2.7
& 6.1 & 41.2[23.5+17.7] & 47.3 \\
\botrule
\end{tabular}
\caption{Partial sums of the contributions of pions and of axial vector mesons to $a_\mu$ in units of $10^{-11}$, where $a_{\mu(L)}$ denotes the longitudinal contribution only. Here $\pi$ and $a_1$ refer to the entire
tower of pions and $a_1$ mesons; $\pi^*$ only to the heavy pions and $a_1^*$ only to excited axial vector mesons. A and P$^*$ refer to a whole $U(N_f=3)$ multiplet of axial vector mesons and excited pseudoscalars, where the contributions from the former are split into longitudinal (L) and transverse (T) parts. In the present flavor-symmetric case 
$a_\mu^\mathrm{A}\equiv 4a_\mu^{a_1}$ and $a_\mu^{\mathrm{P}^*}\equiv 4a_\mu^{\pi^*}$.
}
\label{tab:amutotals}
\end{table}

Table \ref{tab:amutotals} shows the sums of the contributions in the different
sectors, where we have made a numerical estimate of the limit value when
the infinite tower of axial vector mesons is included as in \cite{Leutgeb:2019gbz}; in the case of the
excited pions, the contributions of the higher modes fall off very quickly
so that we have just summed the first few modes. 

For the contributions of the excited pions we obtain the range $a_\mu^{\pi^*}=(0.8\ldots1.8)\times 10^{-11}$ with the
chiral model being at the lower end. Even though we have seen above that the HW models appear to severely overestimate the two-photon coupling $F_{\pi_2\gamma\gamma}$
when comparing with the upper limit \cite{Colangelo:2019uex}  $|F_{\pi(1300)\gamma\gamma}|$ given above, the holographic results are somewhat below
the contributions from the first few excited states obtained in \cite{Colangelo:2019uex} with large-$N_c$ Regge models.\footnote{The model of \cite{Colangelo:2019lpu,Colangelo:2019uex} respects the known experimental constraints, but was constructed such that the longitudinal SDCs are saturated by excited pseudoscalars instead of axial vector mesons.}
In Ref.\cite{Leutgeb:2019zpq} also the contributions from the $\eta$ and $\eta'$
pseudoscalars were estimated on the basis of the chiral HW1 model; we defer a precise evaluation of those in the massive HW models to future work where we plan to study the flavor-asymmetric case together with the contributions from the Witten-Veneziano mechanism for implementing the $U(1)_A$ anomaly. In the present flavor-symmetric setup,
we would simply estimate the contribution of a whole $U(N_f=3)$ multiplet ($P^*$) as $a_\mu^{\mathrm{P}^*}\equiv 4a_\mu^{\pi^*}=(3.4\ldots7.2)\times 10^{-11}$.

The much higher contributions from the infinite tower of axial vector mesons,
which in the chiral HW1 model reads \cite{Leutgeb:2019gbz} $a_\mu^\mathrm{A}=40.6\times 10^{-11}$,
span the range $(39.3\ldots 43.3)\times 10^{-11}$ in the massive HW models, where
the highest value is obtained in the HW1m' model in which the physical mass of the
$a_1$ meson can be fitted by using a nonstandard $M_X$ value. This model has also
the largest contribution from the excited pseudoscalars, so that in combination
$a_\mu^{\mathrm{A}+\mathrm{P}^*}=50.5\times 10^{-11}$ is reached; the lower end
of the results for this quantity is provided by the HW3m model with $43.3 \times 10^{-11}$,
where the masses of axial vector mesons are in fact too high overall.

Generally we find that the contributions from excited axial vector mesons
are more important than excited pions, corresponding to the fact that only
the infinite tower of the former plays a role in satisfying the LSDCs (which
is completely satisfied in the asymmetric Melnikov-Vainshtein case, and
at the level of 81\% in the symmetric case). 
Massive pions already contribute
in the HW1 and HW3 models in the chiral limit; away from the chiral limit
their importance is not increased, despite the
different asymptotic
behavior of their summed contribution in the HW1m and HW3m cases where one has
a logarithmic enhancement
of the TFFs. Correspondingly, the contributions of the axial vector tower are not reduced; in fact, they tend to be higher with nonzero quark masses.

\subsection{Discussion}

\begin{table}[b]
\begin{tabular}{ccccc}
\toprule
$\frac{g_5^2N_c}{12\pi^2}$ & $R(A(0,0)_{n=1})$ & $R(a_\mu^{P_1,P})$ & $R(a_\mu^{\mathrm{A}_1})$ & $R(a_\mu^\mathrm{A})$ \\
\colrule
0.90 & 0.93 & 0.96 & 0.91 & 0.95 \\
0.85 & 0.90 & 0.94 & 0.87 & 0.92 \\
\botrule
\end{tabular}
\caption{Reduction factors $R$ in pseudoscalar and axial vector contributions
to $a_\mu$ as estimated from the chiral HW model when the asymptotic LO pQCD constraints for TFFs are satisfied only at the levels of 90\% or 85\%.
$P$ and $A$ refer to the whole tower of pseudoscalar and axial vector mesons, $P_1$ and $A_1$ to the ground-state modes.}
\label{tab:reductionfactors}
\end{table}

Since we have seen above that the holographic results for the low-energy observables $F_{\pi_2\gamma\gamma}$
and $A_1(0,0)$, the latter determining the equivalent two-photon rate of the lightest axial vector meson, are larger than indicated by experiments, the corresponding holographic results
may perhaps be viewed as upper limits. In the following we investigate whether
this situation improves when one tries to accommodate corrections to the high-energy
behavior.
In fact,
the holographic HW models we have considered here have no running
coupling constant; the TFFs reach their asymptotic UV limits somewhat too quickly.

In order to derive more plausible extrapolations to real QCD,
we have considered a reduction of the value $g_5^2$ by 10\% and by 15\%. This brings
the asymptotic behavior of the TFFs down by amounts that are roughly consistent with perturbative corrections to the leading-order pQCD results at moderately high $Q^2$ values \cite{Melic:2002ij,Bijnens:2021jqo}. At the same time, the right-hand side of (\ref{PiVas}) is increased by a similar amount, which is consistent with the next-to-leading order terms in this expression \cite{Shifman:1978bx}.

With 10\% reduction, the HW model results for the pion TFF also get closer to the central result of the dispersive approach at all energies, while with 15\% they
are generally somewhat lower. In Table \ref{tab:reductionfactors} we have listed the reduction
factors resulting for $A_1(0,0)$ and various $a_\mu$ contributions in the chiral HW1 model which we assume to be a good approximation in general. Applying the stronger reduction factors to the minimum values of the results of the massive HW models, 
we obtain the range 
\bea
a_\mu^{\pi^0}&=&(R(a_\mu^{P_1})_{0.85}\times 64.3\ldots66.6)\times 10^{-11}\nonumber\\
&=&(60.5\ldots66.6)\times 10^{-11},
\eea
in remarkable agreement with recent evaluations using the data-driven dispersive approach \cite{Hoferichter:2018kwz}, where $a_\mu^{\pi^0}=62.6^{+3.0}_{-2.5}\times 10^{-11}$,
which has also been backed up by lattice QCD \cite{Gerardin:2019vio}.
Doing the same for excited pseudoscalars and axial vector mesons, we obtain
the following ranges as our predictions for the contributions of excited pseudoscalars and axial vector mesons:
\bea\label{extendedranges}
&&a_\mu^{\mathrm{P}^*}\equiv 4a_\mu^{\pi^*}=(3.2\ldots7.2)\times 10^{-11}, \nonumber\\
&&a_{\mu(L)}^{\mathrm{A}}=(20.8\ldots 25.0)\times 10^{-11} ,\nonumber\\
&&a_\mu^{\mathrm{A}}=(36.6\ldots 43.3)\times 10^{-11} ,\nonumber\\
&&a_\mu^{\mathrm{A}+\mathrm{P}^*}=(39.8\ldots 50.5)\times 10^{-11}.
\eea

The latter result could be compared to the White Paper \cite{Aoyama:2020ynm} values attributed to the axial sector and contributions related to the SDC,
$a_\mu^\mathrm{WP,axials}=6(6)\times 10^{-11}$ and $a_\mu^\mathrm{WP,SDC}=15(10)\times 10^{-11}$, which with linearly
added errors gives $21(16)\times 10^{-11}$, which is significantly smaller.

In Ref.~\cite{Ludtke:2020moa}, a model-independent estimate of the effects of the longitudinal short-distance constraints on the HLBL contribution to $a_\mu$ has been proposed with the result $\Delta a_\mu^{(3)}=2.6(1.5)\times 10^{-11}$ for the isovector sector.
In Table \ref{tab:amutotals} we have also listed the results for the
longitudinal part of the contributions from the excited pions and the $a_1$ tower. 
Extending the range of the holographic results by the above reduction of the lower end, we obtain 
$a_{\mu(L)}^{\pi^*\cup a_1} =(6.0\ldots 8.1)\times 10^{-11}$
which is significantly higher. However, excluding the ground-state axial vector meson, whose mass is close to the matching scale used in \cite{Ludtke:2020moa} and which (like any single excitation) does not contribute to the asymptotic value of the TFFs, we just have
$a_{\mu(L)}^{\pi^*\cup a_1^*} =(1.7\ldots 3.7)\times 10^{-11}$, in perfect agreement with $\Delta a_\mu^{(3)}$ of Ref.~\cite{Ludtke:2020moa}.
Including singlet and octet contributions, Ref.~\cite{Ludtke:2020moa}
has estimated $\Delta a_\mu^{(0)+(3)+(8)}=9.1(5.0)\times 10^{-11}$,
which is 3.5 times the $\Delta a_\mu^{(3)}$ result. In our flavor-symmetric models, we would have to multiply our results by a factor of 4, giving $(6.9\ldots 15.0)\times 10^{-11}$, which is still agreeing well. It would be interesting to revisit the
additional study in Ref.~\cite{Ludtke:2020moa} where the HW2 results for the
lightest axial vector meson were included and a result was found that exceeded
the contributions of the remaining tower.

\begin{figure}
\includegraphics[width=0.5\textwidth]{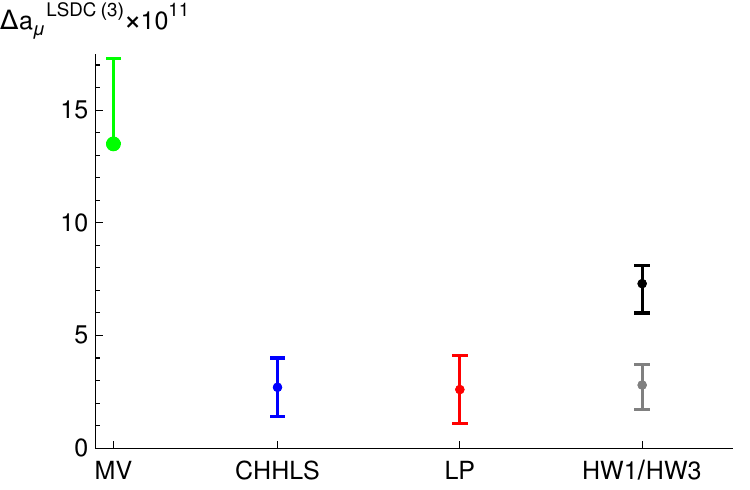}
\caption{Isovector part of the longitudinal contributions beyond the
pion pole. The original estimate of Melnikov-Vainshtein \cite{Melnikov:2003xd} (MV)
of $13.5\times 10^{-11}$ (green dot) translates to $17.5\times 10^{-11}$
with current input according to \cite{Colangelo:2019lpu,Colangelo:2019uex}.
The estimate obtained by the latter,
who have constructed a Regge model for excited pseudoscalars to 
reproduce the MV-SDC,
is shown in blue and labelled CHHLS.
The estimate of L\"udtke and Procura \cite{Ludtke:2020moa} (LP) is plotted in red.
The spread of results obtained in the HW1/HW3 models is shown
in black; the gray version is without the contribution from the
ground-state $a_1$ meson. (The center of the error bar is the center
of the uncorrected results, and the errors are enlarged
downwards according to the 85\% reduction of $g_5^2$.)}
\label{fig:LSDCestimates}
\end{figure}

Another method to estimate the effects of the LSDC has been used in \cite{Colangelo:2019lpu,Colangelo:2019uex}, where a Regge model of excited pseudoscalars has been tuned to reproduce the LSDC. Although we have found in the holographic models that also away from the chiral limit it is the axial vector mesons that are alone responsible for the LSDC, the recently updated estimate obtained in \cite{Colangelo:2021nkr}, $\Delta a_\mu^\mathrm{LSDC}=13(5)$, is fully consistent with our conclusions, as illustrated in Fig.~\ref{fig:LSDCestimates}.

The most important contribution missing in previous evaluations
of the HLBL piece of $a_\mu$, if the holographic HW models
are to be trusted, are those from the ground-state axial vector mesons. With the reduction of $g_5$, also the low energy end
of the axial vector TFF gets modified appreciably, expanding our range of predictions to 
$|A(0,0)|_{n=1}=(17.3\ldots21.3)\,\text{GeV}^{-2}$, which now
has overlap with the experimental values quoted above in Sec.\ \ref{sec:compareTFF}. The lowest value (which also fits the experimental data best) is in fact obtained in the HW1m' model, where the mass of $a_1(1260)$ can be fitted, yielding\footnote{This is significantly larger than our previous ``data-based'' extrapolation in \cite{Leutgeb:2019gbz} which was using an experimental value for $A(0,0)$ that is lower than the ones discussed in Sec.~\ref{sec:compareTFF}.}
$a_\mu^{\mathrm{A}_1}=25.9\times 10^{-11}$.
In this model, however, the $n=2$ axial vector TFF has a larger value than in the other models, and its contribution to $a_\mu$ is also fairly large, so that the second lightest axial vector multiplet
alone contributes another $a_\mu^{\mathrm{A}_2}=7.5\times 10^{-11}$, despite having a mass much higher than that of established excited axial vector mesons. As we have already remarked, the total
contribution of the axial vector tower is the largest of all HW models (see Table \ref{tab:amutotals}). 
With maximal reduction of $g_5^2$ it still yields $a_\mu^{\mathrm{A}}=40\times 10^{-11}$,
coinciding with the central value of the range given in (\ref{extendedranges}).

All in all, the holographic HW models that we have considered here point to $\gtrsim 20\times 10^{-11}$ of extra contributions
in the HLBL part of $a_\mu$ compared to the axial-vector and SDC pieces in the White Paper value \cite{Aoyama:2020ynm} of $a_\mu^\mathrm{HLBL,WP}=92(18)\times 10^{-11}$,
chiefly due to the axial vector meson contributions. Such sizable upwards corrections are
in fact compatible with the recent complete lattice calculation \cite{Chao:2021tvp} with 
comparable errors, which obtained  $a_\mu^\mathrm{HLBL,lattice}=106.8(14.7)\times 10^{-11}$.

\section{Conclusion}

In this paper we have studied various AdS/QCD models with a hard wall and 
with a bifundamental scalar which permits to introduce finite quark masses
and thus to extend our previous work on hadronic light-by-light contributions
in holographic QCD away from the chiral limit.
In the latter it was shown in Ref.~\cite{Leutgeb:2019gbz,Cappiello:2019hwh}
that summation of the contributions of the infinite tower of axial vector mesons 
changes the asymptotic behavior of the HLBL scattering amplitude of individual contributions
precisely such that the Melnikov-Vainshtein LSDC is satisfied. By contrast,
in \cite{Colangelo:2019lpu,Colangelo:2019uex} a Regge model of excited pseudoscalars has been
constructed to achieve the same. Turning on finite quark masses, we found that
in the holographic models the infinite tower of excited pseudoscalars, which
is already present in the chiral limit and in fact has nonvanishing two-photon couplings
despite vanishing decay constants, couples to the axial anomaly and then can
lead to a certain enhancement of their contribution to the asymptotic HLBL amplitude,
but never enough to contribute to the leading terms of the LSDCs. With the standard choice of the
holographic mass of the bifundamental scalar, which determines the scaling dimension
of quark masses and chiral condensates, this enhancement is merely logarithmic;
generalizations are possible, where also power-law enhancements arise, but still below
what is relevant for LSDCs at leading order.

We have also considered the numerical consequences of introducing finite
quark masses on the results obtained previously in the chiral limit, for simplicity
only in the flavor-symmetric limit so that the $\pi^0$ and the $a_1$ sector
can be covered, leaving the $N_f=2+1$ case and also the consideration of
the Witten-Veneziano mechanism for the U(1)$_A$ anomaly to future work.
Doing so we have explored the two sets of boundary conditions that are
possible in hard-wall models, and we have also considered the generalization
of modified scaling dimensions of quark masses and chiral condensates proposed in
\cite{Domenech:2010aq}, which permits to fit either the mass of the first excited
pion or the mass of the lowest axial vector meson.

As displayed in Fig.~\ref{fig:barchart}, the massive HW models lead only to small (positive)
changes in the contributions to $a_\mu$ from the $\pi^0$ and $a^0$ towers compared to the
chiral HW1 model, when in the latter the physical pion mass is inserted manually in the pion
propagator. Compared to experimental data, the two-photon couplings of the lowest axial vector meson,
which is responsible for the second-largest contribution besides the ground state pion,
is somewhat too large, but it becomes consistent with experimental constraints when
the five-dimensional coupling is adjusted such that the LO pQCD values of TFFs is
reduced by amounts corresponding to typical $\alpha_s$ corrections at moderately large energies.
However, also after such adjustments, the contributions from the axial vector and excited pseudoscalar mesons ($\gtrsim 40\times 10^{-11}$)
are significantly larger than in the model calculations that have been used to assess
their role and also the effect of SDCs in the White Paper \cite{Aoyama:2020ynm}, where
$a_\mu^\mathrm{axials+SDC}=21(16)\times 10^{-11}$.

\begin{acknowledgments}
We would like to thank
Hans Bijnens, Luigi Cappiello,
Oscar Cat\`a, Giancarlo D'Ambrosio, Gilberto Colangelo, Franziska Hagelstein, Martin Hoferichter, Jan L\"udtke, Jonas Mager, Massimiliano Procura, and Peter Stoffer for discussions.
J.~L.\ was supported by the FWF doctoral program
Particles \& Interactions, project no. W1252-N27.
\end{acknowledgments}

\appendix

\section{Chiral limit and GOR relation}\label{App:GOR}

In the following we give some details on the chiral limit $M_q\to0$ of the
massive HW models, including the generalization
where $M_X^2$ is allowed to deviate from the standard choice $M_X^2=-3$. The
connection to the chiral HW models with strictly $M_q=0$ is somewhat subtle,
because the weight
function $z/\beta(z)$ in the differential equation (\ref{eq:eomy}) for $y=\partial_z \phi/z$ is
(more) singular at $\zeta\equiv z/z_0 =0$ when $M_q=0$,
\be
\frac{z}{\beta(z)}\sim \left\{
\begin{array}{ll}
   \frac{z_0}{g_5^2 M_q^2} \zeta^{-1+2\alpha}  & \mbox{for}\; \zeta\to0,\, M_q\not=0 \\
   \frac{1}{g_5^2 \Sigma^2 z_0^3} \zeta^{-1-2\alpha} & \mbox{for}\; \zeta\to0,\, M_q=0
\end{array}
\right.
\ee
where $\alpha=\sqrt{4+M_X^2}\in (0,2)$, with $\alpha=1$ for the standard choice $M_X^2=-3$.

The asymptotic behavior of the profile function $y_S(q,z)$ near the boundary $\zeta=0$ reads
\be
y_S(q,z) \sim \left\{
\begin{array}{ll}
\frac{g_5^2M_q^2}{2-2\alpha} \zeta^{2-2\alpha} + a_1 + a_2 \zeta^{4-2\alpha} + \ldots & \mbox{for}\; \alpha>1 \\
g_5^2 M_q^2 \ln\zeta + a_1 + a_2 \zeta^2\ln\zeta + \ldots & \mbox{for}\; \alpha=1 \\
a_0 + \frac{g_5^2M_q^2}{2-2\alpha} \zeta^{2-2\alpha} + a_2 \zeta^2 + \ldots & \mbox{for}\; \alpha<1 
\end{array}
\right.
\ee
and that of normalizable modes is given by
\be
y_n=-g_5 f_{\pi_n}+\left\{
\begin{array}{ll}
c_1 \zeta^{4-2\alpha} +  \ldots & \mbox{for}\; \alpha>1 \\
c_1 \zeta^2  + \ldots & \mbox{for}\; \alpha \le 1 
\end{array}
\right. .
\ee
The mode functions $\phi_n$ and $\pi_n$ vanish at the ultraviolet boundary
according to
\be
\phi_n=-g_5 f_{\pi_n}z^2/2+\ldots,\quad
\pi_n=-\frac{m_n^2 f_{\pi_n} z_0^{2-2\alpha}}{2\alpha g_5 M_q^2}z^{2\alpha}
+\ldots.
\ee

When $M_q\ll \Sigma z_0^2$, the weight function $z/\beta(z)$ is concentrated at small $z$,
where its would-be divergence is cut off by $M_q$.
In this limit, one can approximate the normalization condition (\ref{eq:orthonoy}) by
replacing $y_n^2$ by its boundary value $g_5^2 f_{\pi_n}^2$ and the upper limit of
the integral by infinity, yielding
\be
g_5^2 f_{\pi_n}^2 m_n^2 \int_0^\infty dz \frac{z}{\beta(z)}=1,
\ee
where
\bea
&&g_5^2 \int_0^\infty dz \frac{z}{\beta(z)}=z_0^4 \int_0^\infty d\zeta \frac{\zeta^3}{(z_0M_q \,\zeta^{\Delta^-}+z_0^3 \Sigma\, \zeta^{\Delta^+})^2}\nonumber\\
&&=z_0^4 \int_0^\infty d\zeta \frac{\zeta^{2\alpha-1}}{(z_0M_q+z_0^3 \Sigma\, \zeta^{2\alpha})^2}
=\frac1{2\alpha M_q \Sigma}.
\eea
For sufficiently small $M_q$ we thus obtain
\be
f_{\pi_n}^2 m_n^2 \approx 2\alpha M_q \Sigma.
\ee
For massive pions, i.e., for $n>1$, where $m_n$ approaches a nonzero value
in the chiral limit, this implies that $f_{\pi_n}\to 0$, so they decouple, while the lightest pion with $f_{\pi_1}=f_\pi$ gives rise to the Gell-Mann--Oakes--Renner relation
$f_\pi^2 m_\pi^2=2 M_q \Sigma_q$ for $\alpha=1$, while for $\alpha\not=1$ one
should perhaps rescale $M_q$ and $\Sigma$ before interpreting them as quark mass and
condensate. (The scaling factor mentioned in footnote \ref{footnote:rescaling} drops out here.)

While $y_1$ always satisfies the boundary condition $\frac{z}{\beta}\partial_z y_1=0$ at $z=\epsilon$
with $z/\beta\sim z^{-1+2\alpha}$ as $M_q\to 0$, it does not satisfy such a boundary condition
with $\beta(z)|_{M_q=0}$. From the point of view of the strictly chiral HW model,
$y_1$ corresponds to a solution with the different boundary conditions that pertain to the
one of profile functions $y_S$ (up to an overall factor).
Nevertheless, it can still be normalized by (\ref{eq:orthonoy}), since with the help of the
equations of motion the divergent
integral times the vanishing mass can be recast as
\be
  m_1^2\int_{\epsilon}^{z_{0}}dz  \frac{z}{\beta(z)} y_{1}^2
=  \int_{\epsilon}^{z_{0}}dz\, y_{1}\left[z-\partial_z \frac{z}{\beta(z)} \partial_z \right] y_{1}.
\ee

The holographic wave function of the massless pion can be given in closed form
as the appropriate linear combination of the two Bessel functions
\be
z^{1+\alpha} I_{\pm\frac{1+\alpha}{2+\alpha}}\left(\frac{g_5\Sigma z_0^{1-\alpha}}{2+\alpha} z^{2+\alpha} \right).
\ee
In the special case of the chiral HW1 model with standard $M_X^2=-3$ and thus $\alpha=1$ the
result reads \cite{Grigoryan:2007wn}
\begin{equation}
    y_1=N z^2 \left(- I_{-\frac{2}{3}}(\eta z^3)+\frac{I_{-\frac{2}{3}}(\eta z_0^3)}{I_{\frac{2}{3}}(\eta z_0^3)} I_{\frac{2}{3}}(\eta z^3)\right),
\end{equation}
with $\eta=g_5\Sigma/3$ and
$
    N^2=g_5^2 \Sigma^2 \Gamma(\frac13) \Gamma(\frac23) {I_{\frac{2}{3}}(\eta z_0^3)}/{I_{-\frac{2}{3}}(\eta z_0^3)}.
$

\section{Effects of reducing $g_5^2$}

The following table shows the changes brought about by a reduction of
$g_5^2$ by 10\% (HW1-) and by 15\% (HW1-{-}) in the chiral HW1 model.

\begin{table}[h]
\begin{center}
\begin{tabular}{  l  c c   c  c  c  c  c c c c} 
\toprule
model & PS & $n=1$ & $n=2$ & $n=3$  & AV & $n=1$ & $n=2$ & $n=3$ & $n=4$ & $n=5$ \\
\hline
\hline
HW1- chiral 
& $m_{\pi_n}${\tiny[MeV]} & $0\Rsh135$ & 1849 & 2847 & $m_{\AV_n}${\tiny[MeV]} & 1295 & 2065 & 2944 & 3906 & 4893  \\ 

\cline{2-11} 
$M_X^2 = -3$ 
 & $f_{\pi_n}$ {\tiny[MeV]} & 92.4* & 0 & 0 & $F_{\AV_n}/m_{\AV_n}$  {\tiny[MeV]}  & 166 & 205 & 264 & 311 & 350\\

\cline{2-11} 
 & $F_{\pi_n\gamma\gamma}${\tiny[GeV$^{-1}$]} & 0.274 & -0.209 & 0.161 & $A(0,0)$ {\tiny[GeV$^{-2}$]}  & -19.65 & -4.49 & 0.390 & -2.06 & 0.353  \\ 
 
\cline{2-11} 
& $a_\mu^{\pi_n}\cdot 10^{11}$ & 62.7 & 0.8  & 0.1 & $a_\mu^{\AV_n}\cdot 10^{11}$ & 28.7 & 5.6 & 1.7 & 1.1 & 0.4   \\

\hline
\hline

HW1-{-} chiral 
& $m_{\pi_n}${\tiny[MeV]} & $0\Rsh135$ & 1827 & 2830 & $m_{\AV_n}${\tiny[MeV]} & 1255 & 2027 & 2923 & 3892 & 4882 \\ 

\cline{2-11} 
$M_X^2 = -3$ 
& $f_{\pi_n}$ {\tiny[MeV]} & 92.4* & 0 & 0 & $F_{\AV_n}/m_{\AV_n}$  {\tiny[MeV]}  & 161 & 206 & 265 & 311 & 351 \\

\cline{2-11} 
 & $F_{\pi_n\gamma\gamma}${\tiny[GeV$^{-1}$]} & 0.274 & -0.213 & 0.166 & $A(0,0)$ {\tiny[GeV$^{-2}$]}  & -18.90 & -5.13 & 0.406 & -2.00 & 0.332\\ 

\cline{2-11}

& $a_\mu^{\pi_n}\cdot 10^{11}$ & 61.4 & 0.8 & 0.2 & $a_\mu^{\AV_n}\cdot 10^{11}$ & 27.3 & 5.9 & 1.7 & 1.1 & 0.4\\ 

\botrule
\end{tabular}
\end{center}
\caption{HW1 chiral models with reduced UV asymptotics. HW1- and HW1-{-} have $g^2N_c/12\pi^2=0.90$ and 0.85, respectively.}
\label{tab:reducedhw1}
\end{table}

\raggedright
\bibliographystyle{JHEP}
\bibliography{hlbl}

\end{document}